\documentstyle[11pt]{article}

\def\thefootnote{\fnsymbol{footnote}}
\def\roughly#1{\mathrel{\raise.3ex\hbox{$#1$\kern-.75em%
\lower1ex\hbox{$\sim$}}}}

\def\lsim{\roughly<}
\def\gsim{\roughly>}
\def\be{\begin{eqnarray}}
\def\ee{\end{eqnarray}}
\def\Tr{{\mbox{Tr}}\;}
\def\tr{{\mbox{tr}}}
\def\ben{\begin{enumerate}}
\def\een{\end{enumerate}}
\def\beitem{\begin{itemize}}
\def\eitem{\end{itemize}}

\def\thefootnote{\fnsymbol{footnote}}
\newcommand{\beq}{\begin{eqnarray}}
\newcommand{\eeq}{\end{eqnarray}}
\def\la{\langle}
\def\ra{\rangle}
\def\bi{\begin{itemize}}
\def\ei{\end{itemize}}
\def\ie{{\it i.e}}
\def\eg{{\it e.g.}}
\def\etal{{\it et al}}
\def\del{\partial}
\def\L{{\cal L}}
\def\gV{g_{\mbox{\tiny V}}}

\long\def\beginomit#1\endomit{}
\def\np{{Nucl. Phys.}}
\def\prl{Phys. Rev. Lett.}
\def\pr {Phys. Rev.}
\def\PR {Phys. Repts.}
\def\ijmp{Int. J. Mod. Phys.}
\def\pl{Phys. Lett.}
\def\L{{\cal L}}

\begin{document}


\begin{titlepage}\begin{center}


\hfill{hep-ph/9504250}

\hfill{\it April, 1995}
\vskip 0.4in
{\Large\bf CHIRAL RESTORATION IN}
\vskip 0.1cm
{\Large\bf HOT AND/OR DENSE MATTER}
\vskip 1.2in
{\large  G.E. Brown$^a$\footnote{Supported by the Department of Energy
under Grant No. DE-FG02-88ER 40388} and Mannque Rho$^{b,c}$}\\
\vskip 0.1in
{\large a) \it Department of Physics, State University of New York,} \\
{\large \it Stony Brook, N.Y. 11794, USA. }\\
{\large b) \it Service de Physique Th\'{e}orique, CEA  Saclay}\\
{\large\it 91191 Gif-sur-Yvette Cedex, France}\\
{\large c) \it Institute for Nuclear Theory, University of Washington}\\
{\large \it Seattle, WA 98195, U.S.A.}
\vskip .6in
\vskip .6in

{\bf ABSTRACT}\\ \vskip 0.1in
\begin{quotation}

\noindent Chiral restoration phase transition
in hot and/or dense hadronic matter
is discussed in terms of the BR scaling based on chiral symmetry and
scale anomaly of QCD. The precise connection between the scalar field
that figures in the trace anomaly and the sigma field that figures in
the linear $\sigma$ model is established. It is suggested that in hot and/or
dense medium, the nonlinear $\sigma$ model linearizes with the help of
a dilaton to a linear $\sigma$ model with medium-renormalized constants.
The relevance of Georgi's vector
symmetry and/or Weinberg's ``mended symmetry"
in chiral restoration is pointed out. Some striking consequences for
relativistic heavy-ion collisions and dense matter in compact stars
following stellar collapse are discussed.
\end{quotation}
\end{center}\end{titlepage}


\renewcommand{\thefootnote}{\#\arabic{footnote}}
\section{Introduction}

\indent

One of the most intriguing problems in physics is how chiral symmetry is
restored in matter as   it becomes hot and/or dense. There is presently a
lot of controversy 
and there are many interesting suggestions.

Some time ago \cite{br91}, we proposed that scaling, which is a property of
the Yang-Mills equations at tree level, applied to the dynamically generated
masses of hadrons made up out of the chiral (up and down) quarks. Our
result for vector (V) and scalar ($\sigma$) mesons in the low-energy sector
was
\be
\frac{f_\pi^*}{f_\pi}\approx \frac{m_V^*}{m_V}\approx \frac{m_\sigma^*}
{m_\sigma}\approx\cdots \label{brscaling1}
\ee
The nucleon effective mass scaled somewhat differently \cite{mr88} as
\be
\frac{m_N^*}{m_N}\approx \sqrt{\frac{g_A^*}{g_A}}\frac{f_\pi^*}{f_\pi}.
\label{brscaling2}
\ee
In these equations the asterisk stands for in-medium quantity.

Since our results were obtained by introducing the breaking of scale invariance
with the low-energy chiral Lagrangian through a scalar field denoted
$\chi$, which we called the ``glueball" field, our paper has been
largely interpreted erroneously as tying the scaling (\ref{brscaling1})
and (\ref{brscaling2}) to that of the gluon condensate. We wish to
emphasize that this is not so and clarify where the error in interpretation
is made. In our argument in \cite{br91}, eq.(13), we split the glueball
field into
\be
\chi=\chi_\star+\chi^\prime
\ee
where $\chi_\star$ is the {\it smooth} gluon mean field and $\chi^\prime$
is the fluctuating scalar glueball field. This separation effectively
splits $\chi$ into the  field $\chi_\star$ which scales the quark condensate
and hadron masses, and the ``non-smooth field" $\chi^\prime$
which governs the scaling
of the gluon condensate. In \cite{br91}, the role of the latter was not
clearly specified. The more precise way in which the separation,
guided by some results of lattice calculations, is made
was given by Adami and Brown \cite{adamibrown} who brought out
clearly that there are effectively {\it two} scales in the broken symmetry
sector of QCD \footnote{Several other authors have arrived at different
conclusions. See, {\eg}, \cite{others,pisarski} and footnote \#12.
To distinguish the scaling we are
advocating from those favored by others, we will call ours ``BR scaling" in
this paper.}. These scales can be given in terms of the two ``bag constants":
\ben
\item The bag constant for chiral symmetry restoration
\be
B_{\chi SB}\approx (140 \mbox{MeV})^4.\label{MITB}
\ee
This bag constant is essentially the condensation energy of the vacuum
in the Nambu-Jona-Lasinio (NJL) model \cite{njlvacuum}.
Basically, in this model this is the amount that the condensate of negative
energy constituent quarks is lowered by developing masses. This discussion
is, however, model-dependent and, while the physics seems sensible, it
cannot be satisfactory from a formal point of view. Our aim is to improve on
it.
\item
The bag constant for gluon condensation \cite{shuryak78}
\be
B_{glue}\approx (250 \mbox{MeV})^4.\label{glueB}
\ee
\een

There is an intricate relationship between quark and gluon condensates, as
evidenced by their behavior as the temperature is increased through chiral
restoration $T\sim T_{\chi SR}$. Lattice gauge calculations \cite{kogut91}
show that about half of the gluon condensate decondenses \cite{kochbrown,shlee}
as the temperature rises through $T_{\chi SR}$. Of course, all of the quark
condensate\footnote{We denote the current quark by $q$ and the constituent
quark by $Q$.} $\langle \bar{q}q\rangle$, which can be considered as the
order parameter for the chiral restoration transition, goes to zero
at the critical temperature.

In the chiral limit (defined as the limit in which bare quark masses are
set equal to zero), all of the mass of hadrons such as the nucleon or
vector meson can be expressed in terms of the gluon condensate $T_\mu^\mu$,
since the energy-momentum tensor involves only this condensate. In the
QCD sum rule results, however, the dynamically generated hadron masses depend
chiefly on the quark condensate $\la \bar{q}q\ra$, the gluon condensate
entering as a correction to this, at about the 10\% level.\footnote
{An analogy is that of driving a car. The car moves because the wheels turn
(gluon condensate). However, the driver determines the velocity at
which he is moving from the speedometer (quark condensate). The
wheels and speedometer are technically connected (QCD).}

The intricate interplay between quark and gluon condensates in the
determination of the $\rho$-meson was made clear by Adami and Brown
\cite{adamibrown92}. The argument given there was rather formal, phrased in
temperature-dependent QCD sum rules. We present here a much simpler physical
argument.

The role of the $\la \bar{q}q\ra$ quark condensate is to produce most of the
$\rho$-meson mass. It can be considered as generating the mass by producing
a repulsive scalar potential, equal in magnitude to the mass $m_\rho$,
from the scalar quark condensate in the vacuum, as shown in Fig.\ref{sigma}.
\begin{figure}
\vskip 6cm
\caption[$\rho$ mass] {The role of the quark condensate
 $\la\bar{q}q\ra$ in generating
the $\rho$-meson mass can be visualized in terms of a scalar field coupling
the $\rho$ meson to the negative quark condensate in the vacuum.
This simple picture was developed in \cite{adamibrown91}, in order
to relate the NJL model to this picture of the vacuum condensate producing the
$\rho$ meson mass by this mean-field mechanism.}
\label{sigma}
\end{figure}

Clearly the quark condensate couples repulsively to the $\rho$ meson, the
repulsive scalar potential representing the $\rho$-meson mass. However,
from the sign of the coefficient of gluon condensate, $C_{G^2} (T^2,M^2)$,
this (gluon) condensate couples attractively. The dimension-4 operator $G^2$
where $G_{\mu\nu}$ is the gluon energy-momentum tensor is scaled with
$M_B^{-4}$, where $M_B$ is the Borel mass. With increasing density and
decreasing quark condensate, $M_B\rightarrow M_B^*$, so decreasing.
Although the gluon
condensate enters into determining $m_\rho$ only at the $\sim 10$ \%
level at zero density, at finite density, it would seem to become relatively
more important because of the decreasing $(M_B^*)^{-4}$ effect. The gluon
condensate itself is known to change rather little with density. At the point
where the repulsive contribution from the quark condensate has dropped to the
value of the attractive contribution from the gluon condensate, the
$\rho$-meson mass $m_\rho^*$ would go to zero, before $\la \bar{q}q\ra^*$ goes
to zero. It seems unreasonable that $m_\rho^*$ should go to zero {\it before}
chiral restoration. (There is an exception to this in connection with
Georgi's vector limit which will be discussed in detail below.)

Adami and Brown \cite{adamibrown92} resolved this difficulty by showing that
when the perturbative (black-body) temperature effects are summed and
incorporated into the coefficient $C_{G^2} (T^2,M^2)$, then the
modified coefficient goes to zero as $T\rightarrow T_{\chi SR}$. The
contribution of the gluon condensate to the $\rho$-meson mass is always small,
$\sim 10$ \% of that from the quark condensate.  Pictorially this can be
represented by Fig.\ref{pictorial}. Thus the gluon condensate enters as
a ``Lamb shift" correction to the $\rho$-meson mass. That is, it is the
modification
by the virtual gluon field in the mass dynamically generated from
$\la \bar{q}q\ra$.

\begin{figure}
\vskip 6cm
\caption {Contribution of the gluon condensate to the $\rho$-meson mass.}
\label{pictorial}
\end{figure}

One can thus understand how the dynamically generated $m_\rho^*$ goes to
zero as $\la \bar{q}q\ra^*$ goes to zero. However the scaling
(\ref{brscaling1}) and (\ref{brscaling2}) --
``BR scaling" -- that results from splitting
of $\chi$ into a $\chi_\star$ and $\chi^\prime$, has not been clearly
explained.
A recent paper by Beane and van Kolck \cite{beane} suggests
a remedy to this deficiency.
We describe their development in the next section.

\section{A Dilatation-Invariant Low-Energy Lagrangian}
\indent

The starting point of Beane-van Kolck treatment is an old ``theorem" of
Weinberg \cite{weinberg79} which states that the full content of quantum
field theory (here QCD) is given entirely  by general physical principles
like unitarity, cluster decomposition, Lorentz invariance etc. together
with the assumed internal symmetries. QCD is characterized by the global
$SU(2)_L\times SU(2)_R$ symmetry, which however is not obvious from the
spectrum. Weinberg's theorem suggests that an ``equivalent" field theory
exists, which employs constituent quark ({\it quasiparticle}) fields
that realize chiral
symmetry nonlinearly and therefore includes explicitly Goldstone boson
fields. Brown and Rho \cite{br91} went further, beginning from the
Skyrme Lagrangian out of which the fermions (nucleons) arose as chiral
solitons. This would imply a replacement, at a later stage, of the nucleons
made up out of constituent quarks in the Beane-van Kolck approach by baryon
fields.

The low-energy effective theory that we wish to study must manifest the
same conformal symmetry possessed by QCD. Yang-Mills theory possesses
scale invariance which is broken by the anomaly while the scale invariance
of the fermionic part of QCD is broken by the quark masses. Now the
variables of the low-energy Lagrangian are the Goldstone bosons (pions)
and constituent quarks (or nucleons). Low-energy vector and axial vector
interactions can be viewed as made up out of correlated pions or
in a more compact way introduced as hidden gauge bosons \cite{bando}.
The latter allows us to relate the property of the vector mesons
in hot and/or dense matter to Georgi's vector symmetry \cite{vectorsym}
as we shall discuss later.

The dilatation invariant chiral Lagrangian supplemented by the trace anomaly
that Beane and van Kolck arrive at is
\be
\L&=&\bar{\psi}i(\not\!\!D +\not\!V)\psi +g_A\bar{\psi}\not\!\!A
\gamma_5\psi
-m\chi \bar{\psi}\psi\nonumber\\
&& +\frac 14 f_\pi^2 \Tr (\del_\mu U \del^\mu U^\dagger) \chi^2 +\frac 12
\del_\mu \chi\del^\mu \chi\nonumber\\
&& -\frac 12 {\rm tr} (G_{\mu\nu} G^{\mu\nu}) - V(\chi) +\cdots
\label{beane}
\ee
Here $m$ is the constituent quark mass (which will be defined more precisely
later) and
\be
V_\mu &=& \frac 12 (\xi^\dagger \del_\mu \xi +\xi\del_\mu \xi^\dagger),
\\
A_\mu &=& \frac i2 (\xi^\dagger\del_\mu\xi-\xi\del_\mu\xi^\dagger)
\ee
with
\be
\xi^2=U=e^{i\pi/f_\pi} \ \ \ \ {\rm with} \ \ \pi\equiv {\bf \tau}\cdot
{\bf \pi} (x).
\ee
The trace ``Tr" is over the flavor group and ``tr" over the color group.
Transformation properties of $\xi$ and of the constituent quark doublet
$\psi$ under chiral and conformal transformation are given in \cite{beane}.
What concerns us chiefly here is that a sufficient power of the scalar
field $\chi$ must be put into each term so that the operator scaling
properties, under dilatation, of each term in the Lagrangian  are those of
QCD, {\ie}, that the Lagrangian is dilatation-invariant, except for
the potential term which breaks this invariance. This potential subsumes
radiative corrections of high chiral order and hence can be very complicated,
the only condition being that it gives precisely the trace anomaly of QCD
in terms of the scalar field $\chi$. The precise form of this potential
is neither known nor necessary for our discussion. The basic assumption here
is that the symmetry broken by the anomaly could be represented by that of
spontaneous breaking and that
the Coleman-Weinberg mechanism ``chooses" the
vacuum value of the scalar field {\it in hot and/or dense medium}.

In the ``derivation" of the low-energy effective Lagrangian from a microscopic
Lagrangian (say, QCD or a fundamental modeling of QCD), scaling is
always broken in obtaining the current algebra term $\frac 14 f_\pi^2
\Tr(\del_\mu U\del^\mu U^\dagger)$. Generally, a regularization
is made to incorporate high-energy degrees of freedom in low-energy
effective theories and a momentum cut-off $\Lambda$ that delineates the
high-energy sector from the low-energy one gets transmuted to the physical
constant $f_\pi$, the pion decay constant. The scale breaking through
the potential $V(\chi)$ renders this transmutation automatic and more
or less unique.

In the next section we shall review how Beane and van Kolck relate
the scalar field $\chi$ to an effective scalar $\sigma$ field by
introducing a linear basis
\be
\sigma+i{\bf \tau}\cdot {\bf \pi}=U \chi.
\ee
It is in the linear basis that low-energy current algebra is reconciled
with high energy Regge asymptotics and consequently becomes more relevant
for high-T and high-density processes we are interested in.

As shown in Fig.\ref{sigma} the scalar $\sigma$ field can be thought of as
generating a constituent quark mass $m_Q$ through
\be
m_Q=g_{\sigma QQ} \la \sigma\ra/m_\sigma^2
\ee
where the subscript $Q$ stands for constituent quarks of light flavors.
Compare this to the mean-field model depicted in Fig.\ref{sigma},
\be
\la \sigma \ra=-2 g_{\sigma QQ} \la \bar{q}q\ra/m_\sigma^2,
\ee
the factor 2 coming from the two light-quark flavors.
In terms of the constituent
quark mass $m_Q$ one can then write down the
vacuum energy (see \cite{adamibrown}, eq.(3.11))
\be
E_{vac}&=&-B_{\chi SR}\nonumber\\
&=&- 12\left\{\int_0^\Lambda \frac{d^3k}{(2\pi)^3}
\sqrt{k^2+m_Q^2} -\frac 12\int_0^\Lambda \frac{d^3k}{(2\pi)^3}
\frac{m_Q^2}{\sqrt{k^2+m_Q^2}}\right\} +\frac{3\Lambda^4}{2\pi^2}
\label{evac}
\ee
where $\Lambda$ is a cut-off set to obtain the known empirical value of the
quark condensate
\be
\la 0|\bar{u}u|0 \ra=-(240\ {\rm MeV})^3.\label{qcon}
\ee
Whereas the $E_{vac}$ of (\ref{evac}) is equivalent, physically, to the
condensation energy in the NJL model \cite{njlvacuum} which involves the
quark condensate $\la 0|\bar{u}u|0\ra$ that depends on the
cut-off $\Lambda$ quadratically, eq.(\ref{evac}), however, diverges
only logarithmically with $\Lambda$ which means that the $B_{\chi SR}$
is reliably determined in this formulation.

Determining $\Lambda$ from the quark condensate (\ref{qcon}), one finds
\be
B_{\chi SR}\approx (140\, {\rm MeV})^4\approx 50\, {\rm MeV}/fm^3,
\label{BMIT}
\ee
just the value of the MIT bag constant. We shall see in the next section
that this is just the value obtained from the Beane-van Kolck theory.

We see that in the low-energy sector, dilatation invariance is broken
by the development of the quark condensate. The scale of this breaking
is $(B_{\chi SR})^{1/4}$. In QCD, however, scale breaking occurs in the
gluon sector at QCD loop level. The magnitude of this scale breaking
is an order of magnitude larger than the $B_{\chi SR}$ of eq.(\ref{BMIT});
namely \cite{shuryak78}
\be
B_{glue}\approx 500\, {\rm MeV}/fm^3, \label{Bglue}
\ee
obtained from the QCD trace anomaly. Although $B_{glue}^{1/4}\approx 250$ MeV
is
not so much larger than $B_{\chi SR}^{1/4}\approx 140$ MeV, we know that
the magnitude of the gluon condensate determines the glueball mass
\be
M_{GB}\approx 1.5\, -\, 2 \ \ {\rm GeV}.
\ee
The QCD sum rules tell us \cite{adamibrown92} that the gluon condensate is
relatively unimportant for the masses of hadrons made up out of up and down
quarks, entering only through the Lamb-shift type effect of
Fig.\ref{pictorial}. What this implies is that the scalar $\chi$ field
in the Lagrangian (\ref{beane}) has mostly to do with the quark condensate
\be
B_{\chi SR}\propto \chi^4
\ee
rather than with the gluon condensate $B_{glue}$. Just how $B_{glue}$ reduces
in light-quark systems to $B_{\chi SR}$ is not well understood.
In any event, the $\chi_{\star}$ of
Brown and Rho \cite{br91} should be identified with the mean field of
this {\it  smoothed} field.
This will be discussed in more detail in the next section. Here we consider the
implication of the restoration of the correct scaling properties of the
operators in the low-energy chiral Lagrangian.

The mean-field component of $\chi$ which we denoted $\chi_\star$ previously
will be a function of temperature and density. As in \cite{br91},
we {\it define}
\be
f_\pi^\star\equiv f_\pi \chi_\star.
\ee
We must stress once again
that the $\chi$ field is here already the {\it smoothed} field
of the quark sector that includes {\it no} gluon fluctuation. An equivalent
way of making the separation is to divide the scalar glueball field\footnote{
Instead of dividing the field $\chi$, one could choose to divide
the trace of the energy-momentum tensor $\sim {\rm Tr} G^2$ (where
$G_{\mu\nu}$ is the gluon field tensor) into a ``quarkish" component
$H_q$ and a gluonic component $H_g$ as in \cite{miransky}. This separation
may lead to a different result on scaling, although we have not investigated
this possibility.}
associated with the trace anomaly to a ``smooth" component $\chi_s$
and a ``non-smooth" component $\chi_{ns}$,
$\chi=\chi_{s}+ \chi_{ns}$, and ``integrate" out the non-smooth
component $\chi_{ns}$
from the effective Lagrangian. What appears in the Beane-van Kolck theory is
then $\chi_{s}$ which is relabeled as simply $\chi$.
It is the mean-field component of this field that scales
the quark condensate in the BR scaling.
Its fluctuating part cannot in general describe a single
local degree of freedom since it represents multi-pion excitations including
the continuum but in hot and dense medium can be {\it interpolated} by a local
scalar effective field denoted $\sigma$. We will have more to say on this
point later.

Similarly we define
\be
m_Q^\star\equiv m_Q\chi_\star
\ee
from which we see that the constituent quark mass $m_Q^\star$ scales with
temperature and density in the same way as the order parameter $f_\pi^\star$.
In particular, $m_Q^\star$ must go to zero with chiral restoration,
$f_\pi^\star\rightarrow 0$.

In order to obtain the scaling of the vector meson masses, it is convenient
to look as in \cite{br91} at the quartic Skyrme term
\be
\frac{1}{32 g^2} \Tr\, [U^\dagger \del_\mu U, U^\dagger\del_\nu U]^2.
\ee
One can think of this arising when the vector mesons are integrated out
from the effective Lagrangian. The constant $g$ can be identified as the
hidden-gauge coupling constant. Now this term is scale-invariant as it is,
therefore no $\chi$ field need be multiplied to it. When the
scale invariance is spontaneously broken,
the coefficient of this term remains independent of the factor
$\chi_\star$, so we have
\be
g^\star=g.\label{hgcoupling}
\ee
As shown recently by Harada, Kugo and Yamawaki
\cite{haradaLET}, the effective chiral theory, when
hidden gauge symmetry of the vector mesons is implemented, has
the exact low-energy theorem
\be
m_V^2=2g^2 f_\pi^2\label{ksrf}
\ee
which is essentially the KSRF relation. This theorem
is proven in zero-temperature
and zero-density regime but we see no reason why it should not hold in
medium. This would imply
\be
{m^\star}_V^2=2{g^\star}^2 {f^\star}_\pi^2
\ee
from which, with (\ref{hgcoupling}), follows
\be
\frac{m_V^\star}{m_V}\approx \frac{f_\pi^\star}{f_\pi}.
\label{vectorBR}
\ee

The question of how the nucleon mass scales with temperature and density
is a lot more subtle, because at finite density, a vector mean-field can
develop and build up. There have been several studies on this
issue \cite{Nscaling} but no satisfactory answers have been obtained.
We postpone this question until after we discuss the effective
$\sigma$ field that results once the Lagrangian (\ref{beane}) is transformed
to a linear basis.

We do not discuss the pion mass, which results from explicit chiral
symmetry breaking, a phenomenon associated with the electroweak scale
which is much higher than the chiral scale we are dealing with.
{}From this point of view, the most reasonable thing to do is
to assume that the pion mass does not scale within the range of
temperature and density involved in the chiral Lagrangian.

In \cite{br91}, we deduced the scaling of $\la\bar{q}q\ra$ with
density from the assumed operator transformation properties of the
explicit chiral symmetry breaking term in the low-energy Lagrangian.
The result was
\be
\left[\frac{\la \bar{q}q\ra^\star}{\la\bar{q}q\ra}\right]^{1/3}
\approx \frac{f_\pi^\star}{f_\pi}.\label{suspect}
\ee
For the same reason as for the pion mass, this scaling relation
does not follow immediately (without some strong assumptions) from
the low-energy effective Lagrangian
and in fact may not be quite correct as we shall argue later in connection
with the modeling of lattice results.

It should be remarked that the Brown-Rho consideration was at mean-field
level. The question as to what that corresponds to in the
sense of Weinberg's ``theorem" has not yet been addressed. In a broad
sense of effective theories, the BR scaling is to represent the
tree order in chiral perturbation expansion (more on this later).
Higher-order corrections will surely modify the scaling relation.
For example, from the Goldberger-Treiman relation, we find that
\be
\frac{g_{\pi NN}}{m_N}=\frac{g_A}{f_\pi}.
\ee
As we shall review in the next section, the $g_A$ in medium drops
from 1.26 to $\sim 1$ as density increases from zero to nuclear matter
density $\rho_0$. The cause of this change has to do with the role of
short-ranged interactions between baryons that involve multi-Fermi interactions
absent in the matter-free space. This essentially cancels the change in
$f_\pi$.
Thus $g_{\pi NN}/m_N$ is expected to remain roughly constant up to nuclear
matter density, even though the Brown-Rho scaling would have $m_N^\star$
drop, but $g_{\pi NN}$ remain unchanged. Once $g_A^\star$ has gone to
$\sim 1$, however, one would not expect it to change much further,
and the BR scaling should take over at higher densities.

Before proceeding, we should address the question as to what the precise role
of the effective (density-dependent and temperature-dependent) masses and
coupling constants is in confronting physical observables. As in all field
theory problems, those {\it renormalized}
quantities are {\it parameters} of the theory
that have no physical meaning independently of the
observables one is discussing. For instance, the $m^\star$'s are not by
themselves physical observables. In some special
processes, they could be associated with quasiparticle parameters, that is,
they can be taken as a pole of the particle Green's function but in general,
the pole of the Green's function may not be approximated
by the $m^\star$'s. Residual interactions will modify the residues and pole
positions. In this respect,
it is important to realize that the $m^\star$'s defined
in one theory need not be the same as the $m^\star$'s defined in another
theory. Thus it makes no sense for someone to take a particular hadronic
model to calculate what he/she defines as $m^\star$'s and compare with
another person's $m^\star$'s calculated in a different hadronic model. They
can only compare physical amplitudes computed with the model wherein the
parameters $m^\star$'s appear. This also means that a particular scaling
relation
present at mean-field level of one person's theory need not be reproduced
in another person's mean-field theory. As we shall note below, it is
a particular reparametrization of the fields that gives rise to a simply
scaling theory; other reparametrizations need not give the same scaling
of the parameters although the physics can be identical. In fact, unless one
defines how to compute corrections, the density (and temperature) dependence
of the effective parameters with a particular model has no meaning.
This point seems to be largely overlooked in the field.

This then raises the question: In what sense the BR scaling is to be
understood?

To answer this question, we recall the principal assumption made in
\cite{br91} -- and further elaborated in \cite{mrelaf} --
that at each density and/or temperature, we have an effective
Lagrangian that has the same chiral and conformal symmetry as in free-space
with the parameters of the theory density/temperature dependent.
This can be considered as an application of Nambu's ``theorem" that
whenever spontaneous symmetry breaking
is involved, be that in condensed matter, nuclear
or elementary particles, the physics involved is a generic $\sigma$ model
characterized by different length scales \cite{nambu}.
It is not obvious that this is always possible, so in our case that is
a strong assumption. However we propose
one way of justifying this approach and it is
the notion that the nuclear matter is a chiral Fermi liquid as proposed by
Lynn \cite{lynn}.

It is fairly well-established empirically
that the interactions of mesons and baryons inside nuclear matter
are governed mainly by chiral symmetry. But it is also well-established
that nuclear matter -- and nuclear ground states -- cannot be described
starting from chirally symmetric Lagrangians, at least at low orders.
So the ground-state (mean-field) properties which do not follow naturally
from chiral symmetry and the excitation (fluctuation) properties which do
are not compatible. These can be reconciled, however,
if the nucleus is a soliton, that is,
a chiral Fermi liquid drop with nuclear matter being a chiral Fermi liquid.
The chiral Fermi liquid is a soliton solution of the chiral effective
theory at quantum level.
The parameters so fixed in this soliton structure can be identified with
the $m^\star$'s etc. (In the BR scaling, it is the potential $V (\chi)$
which is supposed to fix the parameters, but we do not really know how
to calculate it. Lynn's chiral liquid approach suggests how to do this.)
Fluctuations around the soliton would then have
the requisite chiral and conformal symmetry of the original Lagrangian,
so must retain the generic $\sigma$ model form. (We argue below that it
is a linear $\sigma$ model.) In looking at the scaling of the $\sigma$
field in terms of phenomenological meson theories that we describe
in the next section, we are implicitly assuming this structure.

While we find the approach described above quite plausible, we must
admit that we have not yet found a realistic chiral liquid soliton
which could be used as the background field to build a theory to
``determine" the parameters in a self-consistent way.
One can however make the following conjecture. To the extent that Walecka's
mean field theory describes nature fairly accurately, the chiral liquid
soliton, when correctly formulated, should resemble Walecka's mean field
theory. The parameters appearing in the effective Lagrangian
describing fluctuations around the chiral liquid (mean field) will
have BR scaling incorporated naturally. Indeed we shall see later
that this is what is observed phenomenologically when we consider
fluctuations in the strangeness flavor, {\ie},\ in kaon-nuclear processes.
Formulating this in a rigorous way remains an open problem.

\section{Transformation to a Linear Basis; Interpretation of the
$\sigma$ Field}
\indent

Beane and van Kolck transform their Lagrangian to a linear
Lagrangian by making the field redefinition
\be
\Sigma=U\chi \label{lineartr}
\ee
where
\be
\Sigma\equiv \sigma +i{\bf \tau}\cdot{\bf \pi}.
\ee
(Note that since the Sugawara field $U=e^{i\pi/f_\pi}$
is conformally invariant, the $\pi$ must scale as $f_\pi$ and
consequently we have a new $\pi^\star$ for each temperature or
density. We shall not make this distinction here.)
The Beane-van Kolck Lagrangian obtained by the above field
redefinition is rather involved, but it simplifies considerably
in the so-called ``dilaton" limit ($m_\sigma\rightarrow 0$,
$g_A\rightarrow 1$) to
\be
\L&=&i\overline{Q}\not\!\!D Q -\frac 12 \tr\big(G_{\mu\nu} G^{\mu\nu}\big) +
\frac 12\del_\mu {\bf \pi}\cdot \del^\mu {\bf \pi}\nonumber\\
&& +\frac 12 \del_\mu\sigma\del^\mu \sigma -\frac{m}{f_\pi}
\overline{Q}[\sigma-i\gamma_5{\bf \tau}\cdot{\bf \pi}] Q\nonumber\\
&&+\frac{m_\sigma^2}{16f_\pi^2}(\sigma^2+{\bf \pi}^2)^2-
\frac{m_\sigma^2}{8f_\pi^2}(\sigma^2+{\bf \pi}^2)^2\ln[(\sigma^2
+{\bf \pi}^2)/f_\pi^2] +\cdots
\ee
Since in the vacuum the pion field has zero VEV and we neglect
pion fluctuation, we can identify
\be
B_{\chi SR}&=&
\frac{m_\sigma^2}{16f_\pi^2} \sigma_0^4=
\frac{m_\sigma^2 f_\pi^2}{16},\label{Bbeane}\\
\sigma_0&\equiv& \la 0|\sigma|0\ra\nonumber
\ee
since $\sigma_0=f_\pi$ in the vacuum.  We find immediately that
\be
B_{\chi SR}^{1/4}\approx \frac 12 \sqrt{m_\sigma f_\pi}.
\ee
For $B_{\chi SR}^{1/4}\approx 140$ MeV, $m_\sigma\approx 918$ MeV, in
the range of $\sigma$ masses in the bare vacuum consistent with
the Weinberg-Tomozawa relation. This $B_{\chi SR}$ is similar to the
condensation energy in the NJL model, eq.(\ref{BMIT}), but has the
advantage that it follows in a straightforward way from the development.
Furthermore, no explicit cut-off is needed in this way of arriving at the
condensate energy, with the bag constant arising naturally at the tree level.
So, although we believe the NJL model captures the essential physics of
the low-energy scale bag constant, the Beane-van Kolck treatment is highly
more preferable and appealing.

Let us now discuss the $\sigma$ field which has emerged in going
from the non-linear
realization of chiral symmetry in the transformation to the linear basis.
We should mention to start with that it is not a mere field redefinition
which would leave physical quantities unchanged. In going from the nonlinear
structure to the linear structure, a nontrivial physics information has
been injected. This point was explained by Adami and Brown \cite{adamibrown}
using a different argument.
The crucial point is that in the transformation, one has
chosen the component of the $\chi$ field such that the $\sigma$ is now composed
mainly of even powers of the pion field $\pi$ in the expansion of the
$U=e^{i\pi/f_\pi}$ field. The fluctuating part of this $\sigma$ therefore
{\it interpolates} two-pion excitations with a mixing to $4\pi$, $6\pi$ etc.
The physical implication of this structure
can be appreciated by recalling the origin of the scalar attraction in the
nucleon-nucleon interaction.

\begin{figure}
\vskip 8cm
\caption {Uncorrelated two-pion exchange involving nucleon and $\Delta$
intermediate states}
\label{2pidiagram1}
\end{figure}

Beginning from the $N\bar{N}\rightarrow 2\pi$ helicity amplitude,
which can be obtained by analytic continuation from $\pi N$ scattering
amplitude, the scalar attraction can be built up \cite{NNpot}.
A dynamical reconstruction of these results produces the scalar exchange
in the Bonn potential \cite{Bonnpot}. In addition to the six uncorrelated
two-pion exchange diagrams shown in Fig.\ref{2pidiagram1}, there are
the contributions in which the two pions scatter on each other in intermediate
states, shown in Fig.\ref{2pidiagram2}.

\begin{figure}
\vskip 8cm
\caption {Correlated two-pion exchange involving nucleon and $\Delta$
intermediate states}
\label{2pidiagram2}
\end{figure}
The $\pi-\pi$ rescattering necessary to form the correlated two-pion
exchange is, of course, automatically included in the dispersion theory
formalism, where the $N\bar{N}\rightarrow 2\pi$ helicity amplitudes are
obtained by analytic continuation of the $\pi N\rightarrow \pi N$
scattering. However it is more convenient for our purpose to use a
dynamical model \cite{dursoetal}
since we shall later introduce effects of finite density and/or temperature.
The correlated two-pion exchange provides about $2/3$ of the total
$2\pi$ exchange in the exchange of scalar degrees of freedom. The mass
distribution is broad, but the correlated exchange can be handled rather
accurately by replacement by a sharp mass particle $\sigma^\prime$.
We shall later drop the prime on this object, because we claim that this is
just the low-mass dilaton of Beane and van Kolck. Effective coupling
constants and masses for this effective scalar particle are given in
various versions of the Bonn potential.

The main rescattering in the Durso-Jackson-Verwest work \cite{dursoetal}
comes from crossed-channel $\rho$-meson exchange. Thus, we can picture
the $\sigma^\prime$ as resulting from the summation of the diagrams shown
in Fig.\ref{sumdiagram}.

\begin{figure}
\vskip 5cm
\caption[scalar] {Description of the scalar meson employed in  terms of
correlated two-pion exchange. The two pions interact chiefly through
$\rho$-meson exchange in the crossed channel.}
\label{sumdiagram}
\end{figure}

We are now ready to investigate the density dependence of the $\sigma$
mass. In eq.(\ref{vectorBR}), we showed that from the KSRF relation
$m_V^\star/m_V\approx f_\pi^\star/f_\pi$; {\ie}, the vector meson mass
drops as the order parameter decreases. This density-dependent $\rho$-meson
mass has been included in the integral equation, shown in Fig.\ref{sumdiagram},
determining the $\sigma$ \cite{dursokim}. From the $N\bar{N}\rightarrow 2\pi$
helicity amplitudes supplied to us by the authors of this work, we can track
the downward movement in mass of the (distributed) scalar strength as the
density increases. Since the downward shift of the scalar strength originates
from the density dependence of the $\rho$-meson mass, it is clear that for
low densities where the term linear in density suffices, that
\be
\frac{m_\sigma^\star}{m_\sigma}\approx \frac{m_V^\star}{m_V}.
\ee

The phase of the $f_+^{J=0} (t)$ helicity amplitude goes through $\pi/2$
for $\sqrt{t}\approx 500$ MeV, showing that the fictitious scalar particle
of mass $\approx 600$ MeV used in boson-exchange models to mimic the
enhancement in the $f_+^{J=0} (t)$ helicity amplitudes for $NN$ scattering
becomes a real resonance of mass $\approx 500$ MeV by $\rho\approx \rho_0$.

Thus, for density $\rho\approx\rho_0$ there is a resonance in the scalar
channel at $m_\sigma\approx 500$ MeV. This light mass $\sigma$
suggests that chiral symmetry be realized in the linear $\sigma$-model as
a manifestation of ``mended symmetry" as emphasized by Beane and van Kolck.
While in the
zero-temperature and zero-density regime, Nature prefers the nonlinear
realization of chiral symmetry for which we have now rather compelling
evidence, as density/temperature is increased in hadronic
matter, the dilaton degree of freedom, frozen in the matter-free vacuum,
gets identified with the effective scalar channel that is interpolated
by a single scalar $\sigma$; at $\rho\approx \rho_0$ the nonlinear
realization of chiral symmetry cedes to a
linear realization with a low-mass $\sigma$
``seen" in boson-exchange models of the nucleon-nucleon interaction.
An important point in our proposition is that in medium, chiral symmetry
is still manifest in Goldstone mode up to the chiral phase transition and
the dynamics can still be described by a chiral Lagrangian. This is in line
with our proposition that Lynn's chiral Fermi liquid (which we tend to
identify with Walecka's mean field) is a background around
which fluctuations be made.

The coupling constants furnish an interesting interplay of the
``vacuum structure" discussed up to now and short-range nuclear correlations
that figure in nuclear many-body systems. At zero density, the pion
coupling constant at tree level is known to be increased by loop
corrections, say, from $g_A=1$ to $g_A=1.26$; {\ie},
\be
g_{\pi NN}=(g_{\pi NN})_{tree} g_A.
\ee
Sometime ago, Rho \cite{mr74} and Ohta and Wakamatsu \cite{ohta}
independently predicted a strong density dependence in $g_A$
\be
\frac{g_A (\rho)}{g_A (0)}&=&\big[1+\frac 89 (\frac{f_{\pi N\Delta}}
{m_\pi})^2\rho \frac{(g_0^\prime)_{N\Delta}}{\omega_R}\big]^{-1}\nonumber\\
&=& \big[1+b\rho/\rho_0\big]^{-1}\label{ROW}
\ee
where $\omega_R\approx 2.1 m_\pi$ is the $\Delta N$ mass difference and
\be
b\approx 0.8 (g_0^\prime)_{N\Delta}.
\ee
Here $(g_0^\prime)_{N\Delta}$ is the Landau-Migdal local field correction
in the channel $NN\leftrightarrow N\Delta$
which can be interpreted as a four-Fermi interaction in the effective
Lagrangian that figures in many-baryon systems \cite{mrelaf}. The
Landau-Migdal parameter is usually taken to be $\sim 1/3$; with this value
we expect\footnote{Careful analyses of Gamow-Teller transitions in nuclei
reveal that the effective $g_A$ in nuclear $\beta$ decay is
(modulo shell effects) near unity
already in light nuclei where density is not high enough for the
Rho-Ohta-Wakamatsu (ROW) mechanism to be fully effective.
This can be understood as follows. In light
nuclei, it is the core polarization effects induced by tensor forces
\cite{arima} that play the main role in
quenching the $g_A$, with the ROW mechanism
being less effective. As density increases in heavy nuclei, the
tensor force gets suppressed by the BR scaling as shown in \cite{brtensor}
and hence becomes less effective in the quenching while the ROW builds up.
In some sense, the tensor correlation and the ROW mechanism play a
complementary
role and both seem related -- {\it albeit} indirectly --
to the chiral restoration phenomenon.}
\be
g_A (\rho_0)\approx 1.
\ee
This rapid drop in $g_A$ means that in nuclear matter calculations, it is
better to use the pion-nucleon coupling given at tree level,
\be
\frac{(g_{\pi NN})_{tree}^2}{4\pi}=\frac{1}{4\pi}(\frac{g_{\pi NN}}{g_A})^2
\approx 8.8.\label{O4const}
\ee
Since the $\sigma$ now forms the fourth component of $SU(2)\times SU(2)\sim
O(4)$, {\ie}, the linear realization of chiral symmetry,
it should have the same
coupling constant (\ref{O4const}). Indeed this is essentially
the coupling of the $\sigma$ in the Bonn potential \cite{machleidt}
where the versions A, B and C of the relativistic OBEP have
$g_\sigma^2/4\pi=8.8$, $8.9$ and $8.6$, respectively.
We thus come to the conclusion that at $\rho\approx \rho_0$, there is
a satisfactory linear realization of chiral symmetry, with $\pi$ and
$\sigma$ mesons coupled with approximately the tree-level
pion coupling to the nucleon. (Note that the in-medium
pion-nucleon vertex in the $\gamma_5$ coupling -- which is what enters
in the linear $\sigma$ model -- is effectively of the form,
$(g_{\pi NN})_{tree}/m_N^\star \approx g_{\pi NN}/m_N$, so as long as
the pion mass does not scale, the pion-exchange interaction remains
unmodified. This feature of the linearly realized chiral symmetry
is consistent with the BR scaling.)
This indicates that nuclear matter
manifests a ``mended symmetry" envisaged by Weinberg \cite{weinbergMS}
in the large $N_c$ limit of QCD.

The density dependences of the type (\ref{ROW}) are normally not included
in the nuclear matter calculations, so it must be somewhat coincidental
that the coupling constant, essentially that of eq.(\ref{O4const}), can be used
at zero density in the Bonn potential to describe the nucleon-nucleon
scattering. At zero density, the low-mass $\sigma$ is not developed, and
the full panoply of uncorrelated and correlated two-pion exchange had to be
worked out. We note that in going from $\rho=\rho_0$ to $\rho=0$ there
are two opposing tendencies. The first is that the resonance in the
correlated two-pion exchange moves up from $\sim 500$ MeV at $\rho\approx
\rho_0$ to $\sim 1$ GeV [{\ie}, $f_0 (975)$] at $\rho=0$.
The second is that the coupling increases
from $(g_{\sigma NN}^2/4\pi)_{tree}\approx 8.88$ to $(g_{\sigma NN}^2/4\pi)
\approx 14$. Even so, Durso, Kim and Wambach \cite{dursokim} find that
the scalar exchange at $\rho=\rho_0$ is substantially more attractive than
that at $\rho=0$. Brown and Machleidt \cite{brownmach} show that
density-dependent loop corrections must be introduced in order to achieve
saturation of nuclear matter. If these higher-order corrections are introduced
into the calculation of $m_\sigma (\rho_0)$, it will not come out as low as
500 MeV. This work is in progress.

The Bonn potential is undergoing some modifications at low momentum scales
resulting from imposition of chiral constraints on the $\pi\pi$ scattering,
which force the off-shell amplitude to go repulsive \cite{aoui}. In an earlier
work, without these constraints, Schuck {\etal} \cite{schucketal} found that
attractive $\pi\pi$ interactions produced very strong effects in the
scalar-isoscalar ($\sigma$-meson) channel, correlating an unreasonable amount
of S-wave strength in the region close to the threshold at $2m_\pi$.
Aouissat {\etal} \cite{aoui} show that when the $\pi\pi$ scattering
amplitudes are constrained by the chiral symmetry \`a la
linear $\sigma$-model \cite{wein67}
the strong near-threshold strength disappears.

In the linear $\sigma$-model, from the potential
\be
V=\frac{\lambda}{4} \left((\sigma^2 +\pi^2)^2-f_\pi^2\right)^2
\ee
the $\sigma$ mass is
\be
m_\sigma^2=2\lambda f_\pi^2.
\ee
We do not expect the constant $\lambda$ to change with density or
temperature, so $m_\sigma^2$ must scale with $f_\pi^2$:
\be
\frac{m_\sigma^\star}{m_\sigma}\approx \frac{f_\pi^\star}{f_\pi}.
\ee
This is the same scaling as obtained in
eq.(\ref{vectorBR}) for the vector mesons.

Once the low-energy $\pi\pi$ scattering is set by the linear $\sigma$-model
\footnote{We are aware of the fact that the linear $\sigma$-model without
matter coupling is not consistent with chiral perturbation theory. Here we
are focusing on low-energy S-wave $\pi\pi$ scattering in the {\it presence} of
matter fields such as nucleons and vector mesons, so that the linear
realization of chiral symmetry does not in practice
suffer from this fundamental defect.},
the introduction of the $\rho$-meson between pions, as discussed above,
must be accompanied by subtractions, so that the S-wave $\pi\pi$ scattering
phase shifts are not destroyed. Built in in this way, the $\rho$-meson plays
a relatively minor role, unlike that in the older work, in the composition of
the effective $\sigma$-meson.

We should mention that
the bag constant $B_{\chi SR}$ of eq.(\ref{BMIT}) is small compared with
$B_{glue}$, but large in comparison with energies in low-energy nuclear
physics. In the nuclear many-body system, it represents an energy of
$\sim 50$ MeV$/fm^3$. At nuclear matter density, the density of the
nucleon is 1/6 per $fm^3$, so that the nucleon rest mass energy is
$\sim 150$ MeV$/fm^3$, and the net binding energy per nucleon of 16 MeV
is only $\sim 3$ MeV$/fm^3$. Thus $B_{\chi SR}$ is actually very appreciable.

%

\section{Chiral Restoration at Finite Temperature}
\indent

Lattice gauge calculations are now sufficiently accurate to make statements
about finite temperature behavior of various quantities, as we discuss.
The situation is not so good with finite density lattice calculations, so
we shall not pursue this matter here.

We first put the finite-temperature lattice results into physical units
following Koch and Brown \cite{kochbrown} who used the results of Kogut
{\etal}\ \cite{kogut91} calculated with 6 time slices. Later, more
extensive results with 8 time slices \cite{gottlieb} are found to
more or less confirm the 6-time-slice results.

In lattice gauge calculations, the quark condensate $\la \bar{q}q\ra$ is
measured for each temperature. In an attempt to understand the lattice
results in terms of effective fields, the entropy was computed in \cite{BBJP}
as a function of temperature, assuming the increase in entropy to result
from the heavier hadrons, $\rho$, $\omega$ and $a_1$,  going massless.
The number of degrees of freedom in hadrons was, however, limited to
24, the number of quarks, since the latter are the fundamental degrees of
freedom. In a modeling of the lattice data along the same line, Koch and
Brown \cite{kochbrown} calculated the rate of increase of entropy in the
quarks under the two scaling assumptions:
\ben
\item That the hadron masses scale to zero as
\be
\frac{m_h^\star}{m_h}=\left[\frac{\la\bar{q}q\ra^\star}
{\la\bar{q}q\ra}\right]^{1/3};\label{scaling1}
\ee
\item That they scale as
\be
\frac{m_h^\star}{m_h}=\frac{\la\bar{q}q\ra^\star}
{\la\bar{q}q\ra}.\label{scaling2}
\ee
\een
The scaling (\ref{scaling1}) is what was obtained in \cite{br91}
with what we now believe to be too naive an assumption\footnote{The
reasoning based on chiral symmetry and trace anomaly that leads to
the BR scaling (\ref{brscaling1}) does not uniquely lead to this relation.
As mentioned below eq.(\ref{suspect}), there is an additional assumption
that goes into it, namely, a relation between
the quark scalar density and the chiral field $U$ which is not precisely
given. As such this relation is not to be identified on the same footing
as eq.(\ref{brscaling1}) although we loosely call it a BR scaling.}
while the scaling (\ref{scaling2}) is obtained in the NJL model for
the scalar field $\sigma$. The NJL model cannot make predictions
for vector mesons whose mass scale is of the same order as the cut-off,
so the model is moot on the $\rho$ and $a_1$.

The scaling  (\ref{scaling2}) which we shall call ``Nambu" scaling
gives an excellent
fit to the lattice results, the lattice entropy increasing with temperature
substantially faster than that calculated from the scaling (\ref{scaling1}).
However, one should be cautious in interpreting the results as a support for
the Nambu scaling. The argument that goes into this theory, namely that
the first 40 (24 quark and 16 gluons) degrees of freedom go massless and
nothing happens to the masses of the other particles, is extremely crude.
It resembles in some sense the Debye theory of phonons, where there is
a sharp cut-off in the phonon spectra at the point where the degrees of
freedom in it equal the number of underlying degrees of freedom.
It must be admitted, however, that crude though it may be, our Debye-like
theory, with linear dependence of the hadron masses on the quark
condensate fits the lattice results for the entropy surprisingly well.
It should be stressed that this fit uses only the linear relation
between the hadron masses and the condensate. The temperature comes in
only in the Boltzmann factor when one calculates the entropy. The temperature
may be somewhat incorrect because Koch and Brown obtained it from the
asymptotic scaling relation, {\ie}, from the lowest-order perturbative
expression. The color coupling constant $\bar{g}$ is not small enough for this
to be quantitatively correct. Thus the temperature scale in Fig.\ref{kbscalar}
may not be accurate. Using a better temperature scale will not, however,
change the fit of the ``Nambu" curve to the lattice data appreciably.

A caveat in what we have been discussing here is that
it might not be a good approximation to have {\it all} the hadron masses
scale universally and hence it would be premature to rule out one option
in favor of the other. What may be considered solid in our argument is
that the lattice results provide an empirical support for the scaling of the
masses together with the connection to the quark condensate as given
by (\ref{scaling2}).

\begin{figure}
\vskip 12cm
\caption[kbs] {The Koch-Brown analytical fit \cite{kochbrown}
to $\la \bar{\psi}\psi\ra$, normalized to unity at $T=0$, is shown by
the dropping dashed line. The data from lattice gauge calculations
\cite{kogut91} is shown without error bars. The effects of the bare quark
masses in ref.\cite{kogut91} were removed  by Koch and Brown, so that their
(dashed) line drops to zero. From the lattice results, the entropy can
be calculated and the result is plotted, again without error bars, as the
rising dashed line. The lower solid line follows from the scaling
(\ref{scaling1}); the upper solid line follows the ``Nambu"
scaling of (\ref{scaling2}).}\label{kbscalar}
\end{figure}

Another interesting information one can get from the lattice results
is the disappearance of the dynamically generated masses of constituent
quarks (or hadrons). Lattice calculations measure  screening masses
of hadrons. DeTar and Kogut \cite{detarkogut} found in four-flavor
calculations that the screening masses of the $\rho$ and $a_1$ mesons
come together, as expected, at $T=T_{\chi SR}$. Furthermore,
to the accuracy measured, the common screening mass came out to be
\be
\omega_{scr}\approx 2\pi T;
\ee
{\ie}, $\pi T$ per quark. Later two-flavor calculations \cite{gottlieb}
confirmed these results and found further that the common screening mass
for the nucleon and its chiral (parity-doublet) partner was
\be
\omega_{scr}\approx 3\pi T
\ee
for $T\geq T_{\chi SR}$. As shown by Adami and Brown \cite{adamibrown},
if a residual mass $m_0$ remained above $T_{\chi SR}$, the screening
mass for the vector mesons would be
\be
\omega_{scr}\approx \sqrt{\pi^2 T^2 + m_0^2}\approx \pi T +\frac 12
\frac{m_0^2}
{\pi T}.
\ee
In fact to the accuracy calculated in lattice gauge simulations, no $m_0$ is
needed, the conclusion being that -- to this accuracy -- $m_0$ is zero.
Of course a small $m_0$ could easily escape detection.

Direct calculation of the constituent quark mass (not the screening mass)
for four flavors \cite{boyd1} yields a mass which drops to
the perturbative value
\be
m_{eff}=gT/\sqrt{6}
\ee
for $T$ going from $T_{\chi SR}$ to 1.75$T_{\chi SR}$. Similar calculations
of the $\rho$-meson mass \cite{boyd2} find $m_\rho^\star$ to decrease
rapidly as $T$ increases. There is of course a caveat to this. One has
to be careful in applying four-flavor results to the two-flavor world,
since the chiral restoration transition is first-order in the four-flavor case
and the quark condensate does not seem to change much below $T_{\chi SR}$,
whereas the transition is a smooth second-order one in the two-flavor case.

Calculations of the Bethe-Salpeter wave functions of $\pi$- and $\rho$-mesons
\cite{bernard} show these mesons to be more compact, {\ie}, more closely
correlated, at 1.5$T_{\chi SR}$ than at zero temperature. Koch {\etal}
\cite{KSBJ} have shown that this can be simply understood
in terms of a dimensional reduction of QCD at high temperature.

The results of ref.\cite{KSBJ} could be summarized as follows.
The quarks propagating space-like experience a space-like potential that rises
linearly at large distances, as first calculated on the lattice by
Manousakis and Polonyi \cite{mansou}. In a ``funny space" obtained by
interchanging $z$ and $t$, the quark-antiquark wave function is
calculated from the non-relativistic Schr\"{o}dinger equation in which
the effective quark mass is
\be
m_{eff}=\pi T.
\ee
As with the screening masses, were a dynamically generated mass still present
above $T_{\chi SR}$, the effective mass to be used for the quark would
be $\sqrt{\pi^2 T^2 +m_0^2}$ where as before $m_0$ is the dynamically
generated mass. Similar calculations of the $\pi$- and $\rho$-meson wave
function were subsequently made by Schramm and Chu \cite{schramm} who extended
them to a wider range of temperature. They found that the wave functions
changed even less with temperature than in the calculation of Koch {\etal}
\cite{KSBJ}.

There are two remarkable features in the theoretical results of Koch
{\etal}: (1) Dimensional reduction of QCD seems to work well right down
to $T_{\chi SR}\approx 140$ MeV, a surprisingly low energy for something
which is supposed to happen at ``asymptotic" energy\footnote{Some insight
into why the effective dimensional reduction works so well has been given
by detailed calculations of Suzhou Huang and Marcello Lissia \cite{huang}.
As noted above,
the lowest Matsubara frequency becomes the chiral mass in our dimensionally
reduced space. Corrections to this dimensional reduction involve only the
higher Matsubara frequencies, the next one being $2\pi T$ higher in
energy. They are prefixed by the running coupling constant $\alpha_s$.
Because of the $2\pi T$, one might expect the scale $\Lambda^2$ to be used in
obtaining this $\alpha_s=\alpha_s (\Lambda^2)$ to be a factor of
$\ln (2\pi)^2$ higher than that for the $\alpha_s$ in the dimensionally
reduced space. The detailed calculation of Huang show it to be  a factor
of $\sim 80$ larger (for $N_c=N_f=3$), so that $\alpha_s$ is cut down by
$(\ln 80)^{-1}
\approx \frac 14$.}; (2) the helicity-zero
state of the $\rho$ meson comes out degenerate with the pion while the helicity
$\pm$ states of the vector meson are degenerate with each other. This is
easily understood with the quarks massless in the dimensional reduction.
In the ``funny space" in which $z$ replaces $T$, at least asymptotically
in temperature, the configuration space with the new $z$ becomes
two-dimensional with only $x$ and $y$ directions. Spin must be either
perpendicular to the ($x,y$) plane or lie in the plane. The $\rho$ meson has
gone massless and behaves like a (charged) photon with helicities
$\pm 1$ perpendicular to the plane. The helicity-zero state which originated
from the longitudinally (in plane) polarized component of the $\rho$
before it went massless now behaves as a scalar and forms a multiplet
with the charged pion. We propose that this situation corresponds to
the Georgi vector limit
\cite{vectorsym} in which the longitudinal components of the massive
$\rho$ meson decouple and become degenerate with the pseudoscalar
Goldstone bosons, the pions, thereby the
$\rho$ becoming massless. Georgi envisaged this limit as arising in
some particular limit (such as large $N_c$ limit) of QCD. It appears from
the results described above that the Georgi vector limit is relevant
in the vicinity of the chiral phase transition. Note that this is also
in line with (though not equivalent to)
Weinberg's ``mended symmetry" scenario exploited by Beane and
van Kolck.\footnote{S. Beane in private communication pointed out that the
Georgi vector limit and Weinberg's mended symmetry are basically different,
particularly near a phase transition.}
We shall have more to say on this matter in connection
with quark number susceptibility in the next section.

%
%
\section{Quark Number Susceptibility; Description of the Vector Interaction
for $T\geq T_{\chi SR}$}
\indent

The quark number susceptibility \cite{gottliebchi,hatsuda} is defined as
\be
\chi_{\pm}=\left(\del/\del \mu_{u} \pm \del/\del \mu_d\right) (\rho_u \pm
\rho_d)
\ee
where the $+$ and $-$ signs define the singlet (isospin zero) and triplet
(isospin one) susceptibilities, $\mu_u$ and $\mu_d$ are the chemical
potentials of the up and down quarks and
\be
\rho_i=\Tr N_i exp\left[-\beta (H-\sum_{i=u,d} \mu_i N_i)\right]/V
\equiv \la\la N_i\ra\ra/V
\ee
with $N_i$ the quark number operator for flavor $i=u.d$.
Generalizing McLerran's
expression \cite{mclerran} for the susceptibility and changing his
baryon number density to quark density,  we obtain
\be
\chi_{\pm}=(VkT)^{-1} \int d^3x \la\la(N_u (x)\pm N_d (x))
(N_u (0)\pm N_d (0))\ra\ra.
\ee
The $\chi_{\pm}$ are called the singlet and triplet susceptibilities.
We shall see that the $\chi_+$ is in the $\omega$-meson channel and
the $\chi_-$ in the $\rho$-meson channel. We can thus call $\chi_{\pm}$
the isosinglet and isovector susceptibilities. The susceptibilities
measured on lattice are given in Fig.\ref{gottliebchi}.
For $SU(2)$ symmetry, we expect $\chi_+=\chi_-$ and this is what one observes
in the lattice results.

\begin{figure}
\vskip 10cm
\caption[gottlieb] {Quark number susceptibilities as calculated by
Gottlieb {\etal} \cite{gottliebchi}. The singlet (isoscalar) susceptibility
is given on the left-hand side and, the non-singlet (isovector) on the
right-hand side. The free-quark susceptibilities $\chi^{(0)}_{S,NS}$, corrected
for effects of the finite lattice size,  are also shown. Horizontal arrows
label the values of $\chi$ in the continuum (where $\chi=0.125$
in lattice units
for 4 time slices) and corrected for the finite size of the lattice (labeled
$8^3\times 4$). The stars on the figure show our values calculated from
perturbative gluon exchange, eq.(\ref{crit1}) and equations that follow.
The star at $T_c$ has to
be moved somewhat to the right, because the phase transition is a smooth one,
of width $\sim 20$ MeV \cite{kochbrown} and our calculation applies only when
the transition is completed.}
\label{gottliebchi}
\end{figure}

Although the singlet susceptibility has large errors, it is statistically
consistent with the more accurate nonsinglet susceptibility. We shall discuss
the latter here. It can be seen that the susceptibility is small at low
temperatures, rising rapidly as $T$ moves through the critical temperature
$T_{\chi SR}$ up to $\sim 70$ \% of the free-quark value, designated
$\chi^{(0)}$ in Fig.\ref{gottliebchi}.

We can distinguish three separate regimes in temperature, which we discuss
one by one. In the very low temperature regime -- in which we are not
particularly interested -- the nonsinglet susceptibility is saturated by the
$\rho$ meson \cite{BBJP}
\be
\chi_{\mbox{\tiny NS}}|_{{\tiny T\ll T_{\chi SR}}}
\approx (kT)^{-1}\int d^3x G_{\rho\rho} (x)
\ee
where $G_{\rho\rho}$ is the $\rho$-meson propagator.

As the temperature $T$ moves upwards to the onset of the phase transition,
the constituent quark model would be more appropriate \cite{kunihiro}.
In RPA approximation of this model as depicted in Fig.\ref{rpa},
the susceptibility below the critical temperature is
\be
\chi=\chi_0/(1+\gV\chi_0)
\ee
where $\gV$ is the coupling of the constituent quark to the vector meson
and $\chi_0$ is the susceptibility for non-interacting quarks which
at $T\approx T_{\chi SR}$ where the dynamical quark mass $m_Q$ has dropped
to zero has the value
\be
\chi_0\approx N_f T^2
\ee
with $N_f$ the number of flavors. Now expressing the constant
$\gV$ in terms of the vector gauge coupling $g$ that figures in the hidden
gauge
symmetry Lagrangian \cite{bando},\footnote{We are not putting asterisks
but the temperature dependence of the constants and masses is understood.}
\be
\gV\approx \frac 14 \frac{g^2}{m_V^2}.\label{gVvalue}
\ee
Thus
\be
\chi (T)\approx \frac{\chi_0 (T)}{1+\frac 14 \frac{g^2}{m_V^2}\chi_0 (T)}.
\label{kunihiro}
\ee

\begin{figure}
\vskip 6cm
\caption[rpa] {The quark number susceptibility below $T_{\chi SR}$ is
described  in RPA approximation by summing quark-antiquark bubbles
interacting by exchange of $\rho$ mesons.}
\label{rpa}
\end{figure}

Kunihiro \cite{kunihiro} investigated in the NJL model
what happens to the susceptibility as expressed in
eq.(\ref{kunihiro}) as temperature approached $T_{\chi SR}$ from below
and concluded that the vector field should decouple to explain the
rapid enhancement of the susceptibility observed in the lattice results.
This means that $g^2/m_V^2$ must steeply go to zero. The NJL model cannot
explain this decoupling since the constant $\gV$ in that model
is a relic of degrees
of freedom integrated out in arriving at the NJL form of effective theory
and hence there is no way to know how this constant runs as a function of
temperature or density. On the other hand the hidden gauge symmetry theory
tells us at least qualitatively how the constant might run.
The hint comes from the recent result of Harada and Yamawaki \cite{haradapl}.

Using the hidden gauge symmetric Lagrangian \cite{bando}, Harada and Yamawaki
showed with one-loop $\beta$ functions that the gauge coupling
$g=2(1+\kappa) g_{\rho\pi\pi}$ (where $\kappa$ is a parameter
which takes the value $\kappa=-1/2$ in free space) scales to zero as
$g(\mu)\sim (\ln \mu)^{-1}$ with $\kappa\rightarrow 0$
when $\mu \rightarrow \infty$. As the gauge coupling goes to
zero, the vector-meson mass which is given by
\be
m_V^2=(1+\kappa)^{-1} f_\pi^2 g^2\label{vecmass}
\ee
goes to zero. We should not take this perturbative argument
too literally for a quantitative understanding
since there could very well
be some important non-perturbative effects in this regime that
could modify the running of the coupling constant but our assumption
is that it is
very possible that the gauge coupling constant drops
-- as suggested by the lattice data -- faster than logarithmically.
When $g=0$, then the vector mesons decouple, the
longitudinal component of the vectors becoming scalar Goldstone bosons
degenerate with the pseudoscalar Goldstone bosons and the vector meson
becomes massless. This is the Georgi vector limit \cite{vectorsym}.
\footnote{The phase with $g\neq 0$ preceding the Georgi vector limit
contains the scalar Goldstone bosons ($s$) that are the
longitudinal components of the massive vector mesons. Since
$\kappa=0$ with $f_s=f_\pi$, the symmetry $SU(2)\times
SU(2)$ is restored for the would-be scalar Goldstone bosons $s$
and the pions. An interesting observation
to make here is that when $\kappa=0$, the direct photon coupling to the
charged pion which is proportional to $(1-\frac{1}{2(1+\kappa)})e$ becomes
$\frac{e}{2}$ whereas in the normal phase where $\kappa=-1/2$ the direct
coupling vanishes. This means that as temperature and/or density is raised,
the photon coupling deviates from the canonical vector dominance picture.
It also means that the photon couples half-and-half directly with
the pion and through the massive $\rho$ for $g\neq 0$, $\kappa=0$ and
with the massless quark-antiquark pair of the $\rho$ quantum number for
$\kappa=g=0$. It should however be pointed out that
as discussed by Pisarski\cite{pisarski}, it is possible to preserve the
vector dominance at all temperatures. This option would then violate
hidden local symmetry with the consequence that the $\rho$ mass will go
up -- instead of down as in BR scaling -- as temperature increases.
While we favor the hidden gauge symmetry prediction which is consistent with
the notion that the vector meson mass is dynamically generated,
nothing rules out
the vector dominance option and it will be up to experiments to
decide which is chosen by nature. This makes the planned dilepton
measurements of the mass shift of the vectors
at GSI, CEBAF and other laboratories particularly
tantalizing.}
In this limit, we will be left with
\be
\chi\approx \chi_0.\label{freechi}
\ee
Before the vector decoupling leading to (\ref{freechi}), we can use the KSRF
relation at $T$ near $T_{\chi SR}$ (which seems to be justified by the
work of Harada, Kugo and Yamawaki \cite{haradaLET}) and $\chi_0\approx 2 T^2$
for $T\approx T_{\chi SR}$ to get the ratio {\it just before the critical
point}
\be
\chi(T^<_{\chi SR})/\chi_) (T^<_{\chi SR})\approx \frac{1}{1+\frac 12\left(
\frac{T_{\chi SR}}{f_\pi}\right)^2}\approx 0.47\label{ratiobefore}
\ee
for $T_{\chi SR}\approx 140$ MeV and $\kappa=0$. Here we are assuming that
as suggested by lattice results,
$f_\pi$ remains at its zero temperature value up to near $T_{\chi SR}$.
(The constant $f_\pi$ is believed to fall very rapidly to zero within a small
range of $\Delta T$ near the critical temperature.) The ratio
(\ref{ratiobefore}) is in agreement
with the lattice data at $T\lsim T_{\chi SR}$.

Let us finally turn to the third regime, namely above
$T_{\chi SR}$.
It has been shown by Prakash and Zahed \cite{prakash} that with increasing
temperature, the susceptibility goes to its perturbative value which can
be calculated with
the perturbative gluon-exchange diagrams of Fig.\ref{perturbation}.
The argument is made with the dimensional reduction at asymptotic temperatures,
but as mentioned above, it seems to apply even at a temperature slightly
above $T_{\chi SR}$.

\begin{figure}
\vskip 6cm
\caption[pert] {Perturbative calculation of $\chi$.
The quark and antiquark coupled to the quark density $\rho_q$ interact
by exchanging gluons depicted by wavy  lines.}
\label{perturbation}
\end{figure}

In the following, we schematize the Prakash-Zahed  argument. Let us assume,
in accordance with
the Georgi vector limit, that the vector meson exchanges are decoupled in
this regime. Above $T_c$, the gluon exchanges are just those of Koch {\etal}
\cite{KSBJ} \footnote{Note that $\sigma_{z,1}\sigma_{z,2}=-1$ for both the
pion and the helicity-zero component of the $\rho$ which is degenerate
with it. This degeneracy of the helicity-zero states should be checked
directly by lattice gauge calculations.}:
\be
V(r_t)=\frac{4\pi e^2}{4m^2}\sigma_{z,1}\sigma_{z,2}\delta (r_t)
\label{V}
\ee
with
\be
e^2\rightarrow \frac 43 \bar{g}^2 T
\ee
for QCD (with $\bar{g}$ the color gauge coupling) and $\delta (r_t)$ is the
$\delta$-function in the two-dimensional reduced space. Here $m=\pi T$ is
the chiral mass of quark or antiquark as explained in \cite{KSBJ}.
Possible constant terms that can contribute to eq.(\ref{V}) will be ignored
as in \cite{KSBJ}.

In order to evaluate the expectation value of the $\delta (r_t)$, we note that
the helicity-zero
$\rho$-meson wave function in two dimensions is well approximated by
\be
\psi_\rho\approx N e^{-r_t/a}
\ee
with $a\approx \frac 23$ fm and the normalization
\be
N^2=\frac{2}{\pi a^2}.
\ee
For the helicity $\pm 1$ $\rho$-mesons, $\sigma_{z,1}\sigma_{z,2}=1$,
so we find that the expectation value of $V$ is
\be
\langle V\rangle=\frac 83 \frac{\bar{g}^2 T}{\pi^2 T^2 a^2}.
\ee

Going from the lowest order process $\chi_0$ to the one involving the first
rung
in the ladder, Fig.\ref{perturbation}, means introducing the correction
factor
\be
1-\frac{\langle V\rangle}{2\pi T},
\ee
the denominator being the unperturbed energy  of the quark-antiquark pair.
Summing the rungs, as shown in Fig.\ref{perturbation}, to all orders gives
us
\be
\frac{\chi}{\chi_0}=\left(1+\frac{\langle V \rangle}{2\pi T}\right)^{-1}.
\ee

The lattice calculations \cite{gottliebchi} use $6/\bar{g}^2=5.32$
which would give $\alpha_s=0.07$ at scale of $a^{-1}$ where $a$ is the lattice
spacing. (The relevant scale may be more like $2\pi/a$.) Calculations
use  4 time slices, so the renormalized $\bar{g}$ is that appropriate
to $a^{-1/4}$. Very roughly we take this into account by multiplying the
above $\alpha_s$ by $\ln 4^2$; therefore using $\alpha_s\cong 0.19$.
With this $\alpha_s$ and the above wave function, we find
\be
\frac{\chi (T_c)}{\chi_0 (T_c)}\approx 0.68.
\label{crit1}
\ee
As can be seen by the $\star$ at $T_c$ on the right-hand graph of
Fig.\ref{gottliebchi}, this is just about the ratio obtained.

In view of the crudeness in our determination of $\alpha_s$, this
quantitative agreement with lattice results may not be taken seriously.
However, it should be noted that the perturbative correction
$\langle V\rangle/2\pi T$ goes approximately as
$T^{-2}$, neglecting the (logarithmic) change in  $\alpha_s$, and
it is seen from Fig.\ref{gottliebchi} that this $T$-dependence
fits that of the calculated $\chi/\chi_0$ quite accurately for
$T >T_{\chi SR}$.

Going towards the vector limit, chiral $SU(2)\times SU(2)$ symmetry
is ``mended," so
\be
f_\pi=f_s\label{f}
\ee
where $f_\pi$ and $f_s$ are the constants defined by
\be
\langle 0| A_\mu^i|\pi^j (q)\rangle =i f_\pi q_\mu \delta^{ij},\ \ \
\langle 0| V_\mu^i|S^j (q)\rangle =i f_s q_\mu \delta^{ij}
\label{axial}
\ee
where $V_\mu^i$ ($A_\mu^i$) is the vector (axial-vector) current.
The isovector scalars $S^i$ correspond to the longitudinal components of the
$\rho$. (We are ignoring here the dilaton $\sigma$ discussed above.)
Thus while for $g\neq 0$, $m_\pi\neq m_S$, the equality (\ref{f})
still holds by the mended symmetry. Now assuming that the matter
above $T_{\chi SR}$ corresponds to the vector limit with
$g=0$, is the relation (\ref{f}) expected to hold ?

As discussed above, in the dimensionally reduced ``funny space,"
the $\pi$ and the helicity-zero $\rho$ are degenerate
(with their dynamical mass equal to zero) and their wave functions
become identical \cite{KSBJ}. Therefore we do expect (\ref{f}) to hold
trivially. This is of course consistent with the picture of the
$\pi$ and $\rho$ made up of two non-interacting massless quarks with the
Matsubara frequency $\pi T$ per each quark. A short calculation
of the matrix element (\ref{axial}) with the $\pi$ wave function in the
``funny space" gives, for large $T$\footnote{We denote the constant
by $\tilde{f}_\pi$
to distinguish it from the physical pion decay constant $f_\pi$.}
\be
\tilde{f}_\pi \sim c\sqrt{\bar{g}} T\label{decay}
\ee
where $c$ is a constant $<< 1$ and $\bar{g}$ is
the color gauge coupling constant. Of course just as the screening mass
has no direct physical meaning -- though perhaps related to physical
quantities
through analytic continuation, one cannot give a physical meaning to
(\ref{decay}) in the sense of the pion decay constant. This may account for
the fact that $\tilde{f}_\pi$ grows with $T$ just as the screening mass does,
while one expects the physical $f_\pi$ to go to zero as one would
for the dynamically generated mass denoted above as $m_0$. We clearly
need to know how to go from ``funny space" quantities to physical space
ones. The link is lacking at the moment. What we can say however is
that the results of the lattice calculations are consistent with
Georgi's vector limit encoded in (\ref{f}).


\section{Effects in Heavy-Ion Collisions}
\indent

Consequences of the vanishing of the hadronic coupling $g_V$ of the hidden
local symmetry, leaving on the colored gluon exchange, are strong for the
relativistic heavy-ion experiments. Firstly, it should be noted that
freeze-out in the Brookhaven AGS experiments has been determined to
be~\cite{AuAu}\footnote{The original determination of $T\gsim 150$ MeV from the
ratio of isobars to nucleons by Brown, Stachel and Welke \cite{BSW}
was corrected about 10 MeV downward by taking effects such as the finite
width of the isobar into account. It should be also mentioned that as recently
shown in a dilated chiral quark model\cite{klr},
the scaling with temperature will be more rapid in dense matter than in
matter-free space. Consequently, the freeze-out temperature in the AGS
experiments where the reaction takes place at a high density must be lower
than the critical temperature determined by lattice calculations
that pertain to zero-density matter.}
\be
T_{fo}\cong 120\sim 140 \ \ {\rm MeV}.
\ee
By freeze-out we mean the effective decoupling, in the sense of energy
exchange, of pions and nucleons. (Less strongly interacting particles, such as
the kaons, freeze out at a higher temperature, say, $T> T_{\chi SR}$.)
Lattice gauge calculations give chiral
restoration at a temperature \cite{lattice}
\be
T_{\chi SR}\cong 140 \ \ {\rm MeV}.
\ee
This suggests that freeze-out for particles other than the pion
and nucleon is at a temperature higher than $T_{\chi SR}$ and that
the pion and nucleon freeze out at about $T_{\chi SR}$.
This means that interactions in the interior of the fireball will be at
temperatures greater than $T_{\chi SR}$.

We have already seen that the hadronic vector interaction essentially
decouples, leaving only perturbative gluon exchange. Hatsuda and Kunihiro
\cite{hatkunihiro} show that the pionic and scalar degrees of freedom
move smoothly through the phase transition and this has been verified
by the behavior of the relevant screening masses as $T$ passes through
$T_{\chi SR}$. This is not surprising since as shown by Wilczek
\cite{wilczek}, the linear $\sigma$ model is the Ginzburg-Landau
effective Lagrangian for the chiral restoration phase transition.
It is in this sense that Beane and van Kolck \cite{beane}
recover the linear
$\sigma$ model in the ``mended symmetry" regime (and also that the
BR scaling arguments make sense at tree level). The pion and $\sigma$ fields
are just the fields of this class of effective theories.

As noted above, the behavior of the pion mass $m_\pi$ at zero chemical
potential
($\mu=0$) is complicated \cite{wilczek}. A common bare quark mass
$m_u\approx m_d=m$ plays the role of an external magnetic
field. Basically one falls back onto the Gell-Mann-Oakes-Renner
relation
\be
m_\pi^2\approx - \frac{2m_q \langle \bar{q}q\rangle}{f_\pi^2}
\ee
extended to finite temperatures and densities. Since both $\langle \bar{q}q
\rangle$ and $f_\pi^2$ go to zero (in the chiral limit) as
$T\rightarrow T_{\chi SR}$, the behavior of the pion mass is a subtle matter.
Indeed for $T=0$ and low densities, one can use the fact that
empirically $m_\pi$ changes only very little (if any) with density,
which we know from the very small (and repulsive) scalar potential from
the nucleons in pionic atoms, so we may turn matters around and say that
$f_\pi$ scales as
\be
f_\pi^\star\propto |\langle \bar{q}q\rangle^\star|^{1/2}
\label{fstar}
\ee
which would keep the pion mass unscaled \cite{weise}.

The AGS and CERN relativistic heavy ion collisions offer exciting new
perspectives on this problem. They make it possible to construct an
environment of high densities, several times nuclear matter density,
{\it and} high temperature, $T > T_{\chi SR}$. As remarked above,
essentially all of times of the fireball existence, which we shall
shortly show to be $\Delta t\sim (25 - 30) \ fm/c$, the temperature
is greater than that for chiral restoration.

\begin{figure}
\vskip 6cm
\caption {The chiral circle.}
\label{chiralcircle}
\end{figure}

\subsection{\it The $\Sigma$-Term Attraction}
Some years ago, Kaplan and Nelson \cite{kaplanelson} showed that
explicit chiral symmetry breaking -- that gives masses to the Goldstone bosons
-- could be ``rotated away" by condensing the Goldstone bosons in dense
nuclear medium. Although they discussed kaon condensation, which we shall
consider later, it is easy to grasp their key idea by looking at the pion.
In fact, the effect we wish to study is interesting for both the pion and
the kaon. Consider the ``chiral circle," Fig.\ref{chiralcircle}.
At finite density, the energy density from the explicit chiral symmetry
breaking
is given by
\be
H_{X\chi SB}=\Sigma_{\pi N} \bar{N}N \cos\theta +\frac 12 f_\pi^2 m_\pi^2
\sin^2\theta.
\ee
The first term on the right-hand side represents the sum of effects
of the explicit chiral symmetry breaking contribution to each nucleon
mass $\Sigma_{\pi N}$ and the second term the sum of the pion mass,
with the pion field being given by $\pi=f_\pi \sin\theta$. Note that
the physical pion results from small fluctuation about $\theta=0$ on the
chiral circle. Expanding $H_{X\chi SR}$ in $\theta$:
\be
H_{X\chi SR}=const. +\frac 12 m_\pi^2 \left(1-
\frac{\Sigma_{\pi N} \bar{N} N}{f_\pi^2 m_\pi^2}\right) f_\pi^2 \theta^2+\cdots
\ee
where the ellipsis stands for higher orders in $\theta$.
We see that the pion has developed an effective mass\footnote{As mentioned
above and to be explained below,
this effective mass is not ``effective" at zero temperature because of
energy-dependent repulsive interactions that enter at the same order of
chiral perturbation theory canceling the sigma-term attraction
but becomes observable as $T\rightarrow T_{\chi SR}$.}
\be
{m_\pi^\star}^2=m_\pi^2 \left(1-\frac{\Sigma_{\pi N} \langle \bar{N}N\rangle}
{f_\pi^2 m_\pi^2}\right).\label{pistar}
\ee
It is, however, known from pionic atoms that the scalar interaction felt by the
pion is very slightly repulsive in nuclei. This repulsion is brought
about \cite{BLRT} by Pauli blocking in the virtual pair terms, of the type
shown in Fig.\ref{pair}. In pion-nucleon scattering, a repulsion
develops because of Pauli blocking when the nucleon in the $N\bar{N}$ bubble
in the pion self-energy tries to go into the state already occupied by the
original nucleon (see Fig.\ref{pair}a). With use of the Wick theorem,
this is described by Fig.\ref{pair}b as a Z-diagram, with the backward-going
line representing the antiparticle.

\begin{figure}
\vskip 6cm
\caption {Pauli blocking.}
\label{pair}
\end{figure}

To fully take into account the effect of this type, one has to include
also the anti-$\Delta$ for pion-nuclear scattering and the anti-decuplet
for kaon-nuclear scattering. The virtual pair terms of Fig.\ref{pair}
and their decuplet counterparts
effectively cancel the attraction expressed in the dropping $m_\pi^\star$,
eq.(\ref{pistar}).\footnote{In the chiral Lagrangian approach of
refs.\cite{BLRT,LJMR,LBR,LBMR}, this repulsive contribution is a part of the
${{\cal O} (Q^2)}$ terms proportional to the square of the meson frequency
that appear as ``counter terms." They are of the same order in the
chiral counting as the $\Sigma$ term of Kaplan and Nelson \cite{kaplanelson}.
This contribution
is important for both pion-nuclear and kaon-nuclear
scattering but as shown in \cite{LBR,LBMR},
it plays an insignificant role for kaon condensation in particular.
The point of the discussion given here and in the following is that as
argued in \cite{LBR,LBMR} this
term can be saturated by what we call ``pair terms" involving the octet and
decuplet whereas in general
or more specifically in chiral Lagrangian approaches
such a term can arise from complicated (uncalculable) sources.}


Up to nuclear matter densities, the $\Delta$-nucleon
energy difference does not change appreciably. In the Appendix of
ref.\cite{BKRW}, this is explained in terms of the local field correction
splitting the nucleon and $\Delta$ at finite density. This correction,
however, depends upon the coupling of vector mesons, $\rho$ and $\omega$,
and as the vector mesons decouple at temperature $T > T_{\chi SR}$,
the local field correction should go to zero. Therefore taking the
decuplet to be degenerate with the octet should be literally correct
at high temperatures.


Our above discussion can be carried over to the case of kaons,
where the effects are larger and the situation is a lot more interesting.


The $K^+$ meson is
\be
|K^+\rangle=|u\bar{s}\rangle
\ee
containing a nonstrange quark and a strange antiquark. The vector interaction
between a $K^+$-meson and the nucleon is, therefore, repulsive and
in dense matter
\be
V_{K^+ N}\cong \frac 13 V_{NN}\cong 90\ {\rm MeV}\,\frac{\rho}{\rho_0}
\cong -V_{K^- N}
\label{repulsion}
\ee
where $\rho_0$ is nuclear matter density and $V_{NN}$ is the vector mean-field
potential felt by a nucleon. The Kaplan-Nelson scalar
attraction, comparable to that giving the pion its effective mass,
eq.(\ref{pistar}), is\footnote{We are using $\Sigma_{KN}\approx 2.83 m_\pi$
obtained at tree level in \cite{LBMR}, but see Appendix.}
\be
S_{K^+ N}\approx -\frac{\Sigma_{KN} \langle \bar{N}N\rangle}
{2 m_K f^2}\cong -64\ {\rm MeV}\,\frac{\rho_s}{\rho_0}=S_{K^- N}
\label{attraction}
\ee
where $\rho_s$ is the scalar density. See also Appendix.
At zero density, virtual pair
corrections, of the type discussed for pions, remove about 37\% of this
attraction \cite{BLRT}.

We note that the chief dependence in the $S_{K^+ N}$  of eq.(\ref{attraction})
is expected to come, at least initially, from that of $f$.
Up to nuclear matter density the decrease in $f^\star$ is
given by eq.(\ref{fstar}) with
\be
\frac{\langle\bar{q}q\rangle^\star}{\langle\bar{q}q\rangle}
\approx 1-\frac{\Sigma_{\pi N}\rho}{f_\pi^2 m_\pi^2}.\label{lineara}
\ee
Now if we take $\Sigma_{\pi N}\approx 46$ MeV, then
\be
{f_\pi^\star}^2\approx 0.6 f_\pi^2\label{ffstar}
\ee
thereby increasing the scalar attraction at $\rho=\rho_0$ by a factor
of $\sim 1.6$.
Whereas the $S_{K^+ N}$ of eq.(\ref{attraction}) is decreased $\sim 37\%$
by virtual pair correction \cite{LBMR}, this factor of $\sim 1.6$ increase
makes up for them, so we believe that the Kaplan-Nelson term (\ref{attraction})
 to be our best estimate for nuclear matter density $\rho_0$.
Extrapolation to higher densities is then made linearly from here.

It should be noted that the (repulsive) virtual pair correction factor
goes as $F\cong (1-.37\omega_K^2/m_K^2)$. In the case of the $K^{-}$-meson,
the $\omega_K^2$ decreases with density, the correction dropping out.
Taking into account the decrease in $(f^\star_\pi)^2$ of (\ref{ffstar}),
we then find that $S_{K^- N}$ becomes roughly equal to $V_{K^- N}$.

\subsection{\it Subthreshold Kaon Production}
Recent experiments on subthreshold kaon production in nucleus-nucleus
collision at $\sim 1$ GeV/nucleon in SIS at GSI \cite{misko}
give  a factor of $\sim 3$ more $K^+$-mesons than predicted by
conventional theories \cite{A} with the best nuclear matter EOS --
one with a conventional compression modulus and momentum dependence,
as we discuss in more detail below.

Implementing (\ref{attraction}) as a correction to the kaon mass, with
$\Sigma_{KN}=2.5 m_\pi$, slightly less than the $\Sigma_{KN}=2.83
m_\pi$ obtained in \cite{LBMR}, Fang {\etal} \cite{B} manage to explain
the enhanced kaon production. Similar results have been obtained by
Maruyama {\etal} \cite{C}. In this second paper, it appears that the
GSI subthreshold $K^+$ events are reproduced without decreasing the
kaon mass with density. But
the authors employ the same scalar mean field for the $\Lambda$-particle
as for the nucleon. However the coupling to the $\Lambda$ should be only
2/3 of that to the nucleon, the scalar field not coupling to the strange
quark. In terms of lowering the in-medium threshold\footnote{Randrup and
Ko \cite{randrup} showed that the subthreshold kaon production was chiefly
determined by the maximum kaon momentum $P_{max}$ which in turn is
determined by the threshold, given the input energy.} for kaon production,
employing the same scalar field for the $\Lambda$ as for the nucleon is,
at least roughly, equivalent to using 2/3 of it for the $\Lambda$ and
1/3 for the kaon. Now 1/3 of the scalar field, for nuclear matter
density $\rho=\rho_0$, is $\gsim 100$ MeV in magnitude,
larger than the Kaplan-Nelson attraction (\ref{attraction})\footnote{
With Brown-Rho scaling and our $\Sigma_{KN}$, the Kaplan-Nelson attraction
will grow up to 80 MeV for $\rho=\rho_0$.}. So the
physics is roughly equivalent.

Fang {\etal} \cite{B} do not include the (repulsive) virtual pair correction,
nor the scaling in $(f^\star_\pi)^2$ of eq.(\ref{ffstar}). Introducing
these for $\rho\sim \rho_0$, they roughly cancel. For $K^+$ mesons,
$\omega_K$ does not change much with density. Initially\cite{LBMR},
it increases
slightly. Thus, the factor $F\cong (1-0.37 \omega_K^2/m_K^2)$ always cuts
the $S_{K^+ N}$ down somewhat. On the other hand, the $(f^\star_\pi)^2$
continues to decrease as $\rho$ exceeds $\rho_0$, so that $\omega_{K^+}$
will come back down to $m_K$ at $\rho\sim (2 - 3)\ \rho_0$, the densities
relevant for subthreshold $K^+$ production. Thus, the scalar attraction and
vector repulsion roughly cancel at these densities. This seems to be
what is required to reproduce the subthreshold $K^+$-mesons\cite{B,C}.

We remind the reader that chiral Lagrangians do not admit a scalar exchange
between the kaon and the nucleon: Scalar fields have no role in chiral
expansion. However, in phenomenological fits to $\overline{K}$-nucleon
scattering \cite{muller}, which do not include the Kaplan-Nelson term,
an attraction of about the same magnitude \cite{BLRT} as would come
from this term is introduced by an effective scalar exchange with
a coupling constant combination
\be
g_{\sigma NN} g_{\sigma KK}/4\pi\approx 0.9. \label{y}
\ee
The $\overline{K}$-nucleon scattering is, however, not so sensitive
to this attraction as is the subthreshold kaon production, where an
attractive scalar interaction substantially increases the production
of kaons.

{}From the above discussion, we see that it should be possible to explain
enough
subthreshold $K^+$ production by using attractive
scalar mean fields for each u or d quark, also the u-quark in the
$K^+$-meson. This is essentially what M\"uller-Groeling
{\etal} \ did,   although we see from eq.(\ref{y}) that the $g_{\sigma NN}
g_{\sigma KK}/4\pi$ they used is somewhat less than the
$\frac 13 (g_{\sigma NN}^2/4\pi)\approx 1.5$--2 usually employed
in nucleon mean fields. Our discussion of Maruyama {\etal} \cite{B} shows that
at least in the determination of $P_{max}$ of the $K^+$, this is essentially
what they did.

Note that the vector mean field should be coupled to each u or d quark, since
it couples to the baryon number; thus each u or d quark experiences a
vector mean field equal to $\frac 13 V_{NN} (\rho)$, as noted earlier.
To the extent that the Kaplan-Nelson term (\ref{attraction}) is equivalent to
the scalar mean field of $\frac 13 S_{NN} (\rho)$ at quark level, we have
the simple picture of the mean fields familiar from, say, Walecka theory,
being applied at (constituent) quark level. The enhanced subthreshold
$K^+$ production then follows from the scalar mean field essentially
canceling the repulsive vector one. The vector mean fields do not affect
the determination of $P_{max}$ of the $K^+$, since the baryon number they
couple to is the same before and after interaction.

Mi\'skoviec {\etal} \cite{misko} make no comparison of their results
with the theoretical results mentioned above, because of uncertainties
in the nuclear matter EOS, and in the $\Delta N\rightarrow K\Lambda N$ and
$\Delta \Delta\rightarrow K\Lambda N$ cross sections, which produce
$\sim 85\%$ of the subthreshold kaons. Questions concerning the EOS
have been sorted out in the past few years \cite{zhang} and the conclusion
is that a conventional compression modulus $K_0=210\pm 30$ MeV \cite{blaizot},
together with momentum dependence (as follows from a nucleon effective mass),
should be used.

\begin{figure}
\vskip 6cm
\caption {Production of $K^+$  and $\Lambda$ in $NN$ and $\Delta N$ collisions
from pion exchange.}
\label{X}
\end{figure}

The elementary processes $\Delta N\rightarrow K\Lambda N$ and $\Delta \Delta
\rightarrow K\Lambda N$ have not been measured experimentally.
However, the $NN\rightarrow K\Lambda N$ cross section is given mainly
by the process of Fig.\ref{X}. The $\Delta$'s that are present act as a
reservoir of energy and the pion in the process of Fig.\ref{X}b is more
nearly on-shell than in Fig.\ref{X}a. The crucial ratio of $\Delta\pi N$ to
$N\pi N$ coupling is known to be $\approx 2$ from analyses of many experiments.
Thus scaling in this way and averaging over the various charge states of the
$\Delta$, as done by Randrup and Ko \cite{randrup}, should be sufficiently
accurate to show that there is a real discrepancy between the experimental
results \cite{misko} and the theoretical ones \cite{A,B,C}.

The $\Delta N\rightarrow NK\Lambda$ and $\Delta\Delta\rightarrow NK\Lambda$
cross sections are being recalculated in J\"ulich \cite{julich}.

The strong interaction calculations, such as those of Ko and collaborators
mentioned above, have been criticized recently by Schaffner {\etal}
\cite{D} and Maruyama {\etal} \cite{E}. Firstly, it is pointed out
that in calculations such as those of Fang {\etal} \cite{B}, the
scalar density $\rho_s=\langle \bar{N}N\rangle$ was ``incorrectly" replaced
by the vector density $\rho_V$. It is clear that the scalar density
decreases relative to the vector density as $m_N^\star$ decreases,
going to zero as $m_N^\star\rightarrow 0$. On the other hand, the ratio
$\langle \bar{N}N\rangle/f^2$ appears in (\ref{attraction}), and
if the vector mass is brought down {\it in medium} by the Kaplan-Nelson
term, then so is $f$ since, as noted earlier, they are related by the
KSRF relation $m_\rho^2=2f^2 g_V^2$ and $g_V$ does not change at tree level.
Thus, the $(f^\star)^{-2}$ would be expected to more than counterbalance the
decrease in $\langle \bar{N}N\rangle$ as compared with $\rho_V$. This
presumption is supported by ref.\cite{C} which mistakenly applied the
same scalar mean field to the $\Lambda$ as to the nucleon. As noted,
for subthreshold $K^+$ production, this amounts to using 1/3 of the
scalar field as the nucleon for the $K^+$; furthermore, this is a somewhat
stronger attraction than would be given by the Kaplan-Nelson term without
scaling the $f^{-2}$.

Schaffner {\etal} \cite{D} further suggest associating the square of the
vector potential with an increase in the kaon effective mass. This
quadratic vector interaction was discarded by Brown {\etal} \cite{F}
on the ground that it meant using, to quadratic order, a term which
had been derived in the chiral expansion only to linear order.
Furthermore these authors argued that effective masses should involve
scalar fields and that the vector mean field should give only a shift
in the chemical potential, as in Walecka mean field theory.

These problems have been simply resolved in the calculation of kaon
condensation in dense stellar matter \cite{BLRT,LJMR,LBR,LBMR}
where heavy-fermion chiral
perturbation theory (HFChPT) was used. In HFChPT, the expansion is made
with the velocity-dependent positive-energy baryon field
$B_v=e^{im_B\gamma\cdot
v}\frac 12 (1+\gamma\cdot v) B$. In \cite{LBMR}, all terms up to
$O(Q^3)$ in the chiral counting are summed. The square of the vector
potential does not enter at this order\footnote{Formally such a
term corresponds to a two-derivative four-Fermi interaction, so that
it would come in at $O(Q^4)$, one order higher than the order calculated
in \cite{LBMR}.}. As explained in a note added in \cite{LBMR}, the
difference $\delta=\rho_s-\rho_V$ appears at $O(Q^4)$ in the chiral
counting and taking into account that difference (and also the
term of the form of the square of the vector mean field) in chiral
perturbation
theory as developed in \cite{LBMR} would require, for consistency,
including $O(Q^4)$ counter terms and calculating leading two-loop
diagrams. A partial account of such terms as is done in \cite{D,E}
would violate chiral Ward identities. We claim that the result of \cite{LBMR}
corroborates this point: The large effects claimed by Schaffner {\etal}
\cite{A} and Maruyama {\etal} \cite{B} do not appear, in that the scalar
field here plays a relatively minor role. Its magnitude is varied
more than a factor of two between calculations in \cite{BLRT} and
\cite{LBMR}, but the results are not so different. It thus appears that
the schematization of \cite{A} and \cite{B} to relativistic mean field
theory is too crude an approximation. We also point out that the
argument of Schaffner {\etal} \cite{A} for repulsion arising from
going off-shell in the interaction is incorrect. Such repulsions
do not exist. As demonstrated in Appendix F of \cite{LBMR}, the physics
does not depend on the way the kaon field interpolates.

Schaffner {\etal} \cite{D} point out the inconsistency of using the chiral
vector and scalar interactions for the $K^{-}$ mean field in the Walecka-type
mean-field formalism for the nucleons (or nucleons and hyperons).
The chiral Lagrangian gives $V_{K^+ N}\cong \frac 16 V_{NN}$ rather than
the $\frac 13 V_{NN}$ we discussed in eq. (\ref{repulsion}). In Sakurai's
work and in the schemes treating the vector mesons as gauge particles in an
effective theory, which we discussed in Section 5, the $\omega$-meson
should couple to the baryon number lodged in nonstrange quarks. The
$\rho$-meson couples to the isospin of the nucleon and this, plus the
coupling to the pion, is responsible for the Weinberg-Tomozawa term in
pion-nucleon scattering. The universal vector coupling can be obtained
from the KSRF relation (\ref{ksrf}), and from it, the $\rho NN$-coupling:
\be
\frac{(g/2)^2}{4\pi}=\frac{g_{\rho NN}^2}{4\pi} = 0.70 (5),
\ee
the $(g/2)$ factor entering here because the isospin of the nucleon is $1/2$.
Via $SU(3)$ symmetry, which is incorporated in the chiral Lagrangian, we
obtain for the $\omega NN$ coupling
\be
\frac{g_{\omega NN}^2}{4\pi}=9\frac{g_{\rho NN}^2}{4\pi}\approx
6.34.\label{83b}
\ee
This is substantially smaller than the $\omega NN$ coupling constant
employed in Walecka-type mean-field calculations:
\be
\frac{g_{\omega NN}^2}{4\pi}\cong 10\sim 12. \label{walecka}
\ee
(The chiral $\omega NN$ coupling of (\ref{83b}) would give a vector mean
field of only 178 MeV at nuclear matter density.) For the chiral calculation,
we are justified in using the density dependence (\ref{ffstar}) for $f_\pi$
derived from chiral considerations. Inserting (\ref{ffstar}) with the KSRF
relation, we find
\be
\frac{g^2_{\omega NN} (\rho=\rho_0)}{4\pi}\approx 10.6.\label{chptbr}
\ee
In other words, the connection between the chiral mean fields and those used
in Walecka mean-field theory, which generally works at $\rho\approx \rho_0$,
can be made {\it if and only if} BR scaling is taken into account.
Neither theory employs form factors, so that comparison of coupling constants
is straightforward.

Calculations to date of kaon condensation in the Walecka mean field formalism
have, indeed, been inconsistent in that they use empirical mean fields in
baryonic interactions, but scalar and vector mean fields for the kaon
from chiral Lagrangians without scaling $f_\pi^\star$.

Our argument leading to (\ref{ffstar}) is difficult to employ for densities
exceeding $\rho_0$. Even if the pion mass remains essentially unchanged,
so that we can employ (\ref{fstar}), the linear extrapolation (\ref{lineara})
cannot be extended to higher densities. On the other hand the relativistic
VUU calculations which employ Walecka mean fields work well up to several
times nuclear matter density.

We can turn matters around, and use the Walecka mean fields determined
by the VUU calculations. But then the $V_{K^+ N}$ should be obtained
as in (\ref{repulsion}), by the scaling indicated by quark counting, from
$V_{NN}$.

What to do with the Kaplan-Nelson scalar term is more intricate, since its
orgin is in the chiral symmetry breaking part of the chiral Lagrangian.
The factor $F\cong (1-0.37 \omega_K^2/m_K^2)$ which arises from virtual
pair corrections at the same order in chiral counting as the Kaplan-Nelson
term, must be included in an ad hoc fashion in the relativistic mean field
formalism. Aside from this factor, scaling (\ref{attraction}) by the
factor $(0.6)^{-1}$ of (\ref{ffstar}) gives $S_{KN}\approx -107$ MeV and
our determination of $S_{KN}$ from the lattice calculation of
$\la N|\bar{s}s|N\ra$, as described in the Appendix, would give
$S_{KN}\approx -120$ MeV. Assuming quark scaling, this would give
$S_{NN}\approx -360$ MeV for the nucleon mean field, not far from that in
Walecka theory.

Although the scalar attraction on the kaon does not arise from a Walecka-type
mean field mechanism, since the kaon is a Goldstone boson, use of the Walecka
theory would give a rather similar effective scalar mean field.

The factor $F\cong (1-0.37 \omega_K^2/m_K^2)$ is, however, important.
For the $K^+$-meson for which the $\omega_{K^+}$ increases above $m_K$
for low densities at least, this factor cuts down the scalar attraction and
makes the overall mean field repulsive, as is observed in $K^+$-nucleus
scattering. In the case of $K^-$-mesons, this factor moves towards
unity as $\omega_K$ decreases with increasing density. Thus, it is not
very important in determining the critical density for kaon condensation
and the equation of state in the kaon-condensed phase.

The importance of the above development is that it enables us to directly use
information obtained from the Bevalac and SIS heavy-ion experiments, as
described by tbe relativistic VUU transport in the calculation of kaon
condensation. Thus we have empirical determinations up to $\rho\sim 3\rho_0$.
As noted, from the subthreshold $K^+$ production experiments, we can
already say that the vector and scalar mean fields, the latter corrected
by the factor $(1-\omega_{K^+}^2/m_K^2)$, are roughly equal at
$\rho\sim (2\frac 12 - 3)\,\rho_0$.

Within the Walecka-type mean field description, we can now understand
the large attraction, $-200\pm 20$ MeV at $\rho=0.97 \rho_0$, for
$K^-$-mesons, found by Friedman, Gal and Batty\cite{friedmanetal}.
This is just the sum of scalar and vector mean fields for the $K^-$.
The number should not be surprising, since the same contribution
of mean fields comes in the spin-orbit interaction
for the nucleon in Walecka theory, and the above interaction
for the $K^-$ would imply that $S_{NN}+V_{NN}\cong 600$ MeV which is
big, though somewhat smaller than needed for agreement with experiments.
(Of course we should not forget that our virtual pair correction cuts
down the $K^-$ interaction somewhat, $\sim 10\%$ at $\rho\sim \rho_0$.)


\subsection{\it ``Cool" Kaons}
The $14.6$ GeV $^{28}{\rm Si} + {\rm Pb}\rightarrow K^+ (K^-) + X$
preliminary data \cite{stachel} show cool components, with effective
temperature of 12 MeV for $K^+$ and 10 MeV for $K^-$, which cannot be
reproduced in the conventional scenarios employed in event generators.
The latter give kaons of effective temperature $\sim 150$ MeV. It
is clear that some cooling mechanism is necessary to produce the
cool kaon component. It is also clear that the above vector repulsion
must essentially be absent for the temperature relevant for these
experiments, as we discuss.

In an interesting article, V. Koch \cite{koch} has shown that
given the attraction of (\ref{attraction}) -- although his results are
not sensitive to the precise amount of attraction as long as it is as
large as (\ref{attraction}) --a cool kaon component can be reproduced.
See Fig.\ref{koch}.
Aside from the attractive interaction, it is necessary that the fireball
expand slowly. The slow expansion results because the pressure in the region
for some distance above $T_{\chi SR}$ is very low \cite{kochbrown}, the
energy in the system going into decondensing gluons rather than giving
pressure.
This results in an expansion velocity of $v/c\sim 0.1$.
In the case of 15 GeV/N Si on Pb transitions, the fireball has been measured
\cite{braun} through Hanbury-Brown-Twiss correlations of the pions to increase
from a transverse size of $R_T (Si)=2.5$ fm to $R_T=6.7$ fm, nearly a factor
of 3, before pions freeze out. With an expansion velocity of $v/c\sim 0.1$,
this means an expansion time of $\sim 25 - 30$ fm/c. (The full expansion
time cannot be measured from the pions which occur as a short flash at the
end.) Thus the fireball expansion lasts a very long time.

As long as the expansion is slow, the adiabatic invariant
\be
I=\int {\bf p}\cdot d{\bf r}\label{adiabatic}
\ee
remains a useful concept for the kaons. If the radius of transverse
expansion $r$ increases by a factor of 3, then the momentum will
decrease by the same factor, and the energy by a factor of $\sim 9$.
Thus, as long as the expansion is slow, the kaons will be greatly cooled,
say, by the factor of $\sim 9$.

\begin{figure}
\vskip 15cm
\caption[ent] {Entropy, energy and pressure densities deduced by Koch and
Brown \cite{kochbrown} from the lattice gauge calculations of Kogut {\etal}
\cite{kogut91}.}
\label{entropy}
\end{figure}

Before proceeding further, we pause here to elaborate on the slow-expansion
scenario for which the low pressure is essential. In  Fig. \ref{entropy}
we give the pressure and energy densities, along with the entropy, deduced
by Koch and Brown \cite{kochbrown} from Kogut {\etal}'s \ lattice
simulations \cite{kogut91}. Strictly speaking, the energy and pressure
cannot be split up into a part from the quarks and a part from the gluons,
for such a division is not gauge-invariant. So we should consider only the
sum.

What we see from Fig. \ref{entropy} is that the pressure is only slightly above
zero in the region where the entropy increases rapidly, {\ie}, in the
region of the phase transition at $T_c\sim 140$ MeV. As noted in
\cite{kochbrown}, this small pressure results because the energy chiefly goes
into decondensing the quarks and gluons, thereby producing almost no pressure.
This can be expressed in terms of an effective bag constant $B$ for the
transition. In order to see how this goes, consider the simplified
case of a transition from pions to a quark/gluon plasma \cite{BBJP}.
For simplicity, we neglect the pion pressure and energy, which can easily
be corrected for at the end. The energy-density of a quark/gluon plasma is
\be
\epsilon_{QG}&=& \frac{37}{30} \pi^2 T^4 + B_{eff},\nonumber\\
P_{QG}&=& \frac{37}{90}\pi^2 T^4 -B_{eff}.\label{89}
\ee
Now the phase transition occurs as soon as the $P_{QG}$ can be brought
positive. From lattice calculations, we know that $T_c\approx 140$ MeV.
{}From $P_{QG}=0$ at $T_c$ we find
\be
B_{eff}^{1/4}=199 {\rm MeV},
\ee
midway between $B_{\chi SR}$ of eq.(\ref{MITB}) and $B_{glue}$ of
eq.(\ref{glueB}). Koch and Brown show that this ``half-way house" results
because only $\sim \frac 12$ of the gluon condensate is ``melted" across the
transition region. Note that if we were to calculate $T_c$ from first
principles, we would need to know this latter fact. In other words,
the determination of $B_{eff}$, that fraction of $B$ which is melted, must
first be found. By using the lattice results, we have turned the problem
around, obtaining $B_{eff}$. Equation (\ref{89}) makes it clear why the
pressure is so small in the region of $T_c$.

The smooth transition that we are finding implies that instead of having
the ``melting" occurring at a fixed $T=T_c$, it occurs over a region,
determined as $\Delta T\sim 20$ MeV by Koch and Brown \cite{kochbrown},
although
this number may change as nonperturbative effects are included in the
$\beta$ function.

Estimates of the energy densities reached in the Brookhaven AGS collisions are
quite model-dependent, but generally indicate that the highest temperatures are
not more than $\sim 170$ MeV, and probably are less than this. Consequently,
the systems do not reach temperatures much above $T_c$.

We should remark that while the relationship between hadron masses and the
quark condensate, investigated in \cite{kochbrown} does not involve
knowing the temperature, the width of the phase transition ($\sim 20$ MeV)
does depend upon the asymptotic scaling relation being {\it effectively} valid.
One can approximately check the validity of this relation as follows:
Asymptotic scaling can be expressed as
\be
\frac{8\pi^2}{\bar{g}^2}=(11-\frac 23 N_F) \ln \frac{k}{\Lambda_{QCD}}
\label{ascaling}
\ee
giving the relation between the color gauge coupling
$\bar{g}$ and the momentum $k$. For a lattice
of $6\times 12^3$, Kogut {\etal} \cite{kogut91} found the transition
at $\beta_c=6/g_c^2\approx 5.34$, whereas for $8\times 16^3$, Gottlieb
{\etal} \cite{gottlieb} find $\beta_c=5.54$. Assuming the scale to depend
chiefly on the number of time slices, one can get rid of $\Lambda_{QCD}$
using the two simulations, in order to express
\be
\delta\beta = \frac{6}{8\pi^2} (11-\frac 23 n_F) \ln (\frac 86)\approx .21.
\ee
This ``theoretical" $\delta\beta_c$, which assumes asymptotic scaling, is to
be compared with the $\delta\beta_c
\approx 0.2$ found in the lattice simulations.
The good agreement indicates that the asymptotic scaling relation should
be adequate for determining temperature differences.

Now in expansion of the system following maximum temperature, the expansion
will ``stick" for a long time (estimated to be 25 -- 30 fm/c) just in the
region of $T_c$, where the pressure is nearly zero. Entropy density is
decreased not by expansion, but by the hadrons, which are formed in this
region, going back on-shell and  the number of heavy hadrons then
decreasing because of small Boltzmann factor $e^{-m_H/T}$. This picture seems
to be supported by the work of Shuryak and Xiong \cite{SX} who find evidence in
the excess photons and dileptons in the SPS collisions (200 GeV/nucleon
collision at CERN) for a long ``mixed" phase of $\tau\sim 30$ -- 40 fm/c.

\begin{figure}
\vskip 15cm
\caption[vkoch] {\small  Kaon spectrum from a full transport calculation
for 14.6A GeV/c Si + Pb collisions\cite{stachel}. The full line is the result
including  mean fields for baryons and kaons, the dotted line the result
for kaon mean field and no mean field for the baryons and the dashed line
the pure cascade result. The calculated results are in arbitrary units. The
nucleon vector and scalar couplings of the mean field were taken to be
$g_V=5.5$ and $g_S=9.27$, the vector coupling being about half of the
Walecka mean-field model, so that the kaons -- which felt 1/3 of the
nucleon mean fields -- experience  an attractive potential of
$U_0\simeq 50$ MeV. The field energy $\frac 12 m_S \phi_S^2$,
where $\phi_S$ is the scalar field,  plays the role of the bag constant
(\ref{MITB}). Its value, taken to be about half of the $B$ of
(\ref{glueB}), was adjusted so that the pressure was low in the region of
the phase transition, in accordance with the lattice results of
Fig.\ref{entropy}.}
\label{koch}
\end{figure}


Now returning to the problem of the cool kaons, the important point
to note is that the kaons move through chirally restored
matter, in which the hadronic (as opposed to gluonic) vector interactions
are small or zero during the $25 - 30$ fm/c period. Kaons which do collide
strongly inelastically will be removed from the cool kaon component
and join the thermal kaons.


Given the cool kaon results, the collaboration E814 -- now E877 --
carried out a study of this component in Au+Au collisions at 10.8 A GeV/c at
the
AGS\cite{AuAu}. Both $K^{\pm}$ spectra exhibited a cool, nonthermal component
but with apparent temperatures in the ranges of $\sim 50$ -- 100 MeV
(depending on the rapidity), substantially higher than $\sim 10$ --
12 MeV found for Si+Au. This higher effective temperature was predicted
by Koch\cite{koch}. The Au system is larger, and the kaons have more
chance to hit something and be scattered out of the cool component.
What was unexpected was that, with decreasing $P_t$, the spectra began to drop
in the case of the $K^+$ spectrum but not for the $K^-$ one. (Following
the presentation of the work\cite{AuAu} at the Quark Matter '95 meeting,
C.Y. Wong suggested that the drop in $K^+$ spectrum with $P_t$ decreasing
below 140 MeV was due to the Coulomb interaction which would be
substantial in the large Au system. Work by Koch, in a schematic
model\label{koch95}, showed this to be a plausible explanation.
Further work with full transport by Koch is in  progress.)

It can be seen that the discovery of the cool kaon component tells us
that the main part of the system is chirally restored (in the form of
the Georgi vector limit), for a long time of $\sim 25 - 30$ fm/c, until
the kaons freeze out. They freeze out well ahead of the pions, which
equilibrate with the nucleons down to a density, of $\rho_{fo}\simeq
0.38\rho_0$, because of their weak interaction.

Since masses of particles go back on-shell in free space
before the particles reach the detector, the bulk of experiments is insensitive
to what the masses were inside the hot and dense medium. Only carefully
designed experiments will tell us that, and the cool kaon component is the
most important of these to date.

\section{Discussion of Chiral Restoration with Temperature}
\indent

We have seen in the last section that the behavior of the quark number
susceptibility with temperature, as calculated in lattice gauge calculations,
shows that the hadronic vector coupling disappears as $T$ moves upwards
through $T_{\chi SR}$, and that the perturbative color gluon exchange
describes the susceptibility well above $T_{\chi SR}$, as argued by
Prakash and Zahed \cite{prakash}. Somewhat surprising is the fact that
the perturbative description, which gives a $1/T^2$ behavior in the difference
$\chi (T)-\chi (0)$ between the susceptibility and that for free quarks,
sets in just above $T_{\chi SR}$; {\ie}, for this purpose, asymptotia is
$T \gsim T_{\chi SR}$. Note that the screening mass of the
$\rho$-meson goes to $2\pi T$, its asymptotic value, as soon as $T$ reaches
$T_{\chi SR}$.


As noted preceding eq.(\ref{vecmass}), the hidden local gauge coupling $g$,
at one loop order, scales to zero as $g(\mu)\sim (\ln\mu)^{-1}$, {\ie}, the
Georgi vector limit.  As shown by Harada and Yamawaki \cite{haradapl},
there is
an ultraviolet fixed point $\kappa=0$. Of course, the one-loop calculations
are most certainly unreliable for learning what happens near the phase
transition, where the higher loop effects and nonperturbative effects
presumably pile up. Even so, the qualitative picture suggested by this
one-loop calculation appears to be correct.

It may be that the increase in $f_\pi^\star=f_s^\star$ as $T$ goes through
$T_{\chi SR}$ found in the last section is an artifact of the spacelike
propagated wave function. With timelike propagation, which has not been up
to now  carried out on the lattice, $f_\pi^\star$ and $f_s^\star$
may go to zero as $T\rightarrow T_{\chi SR}$, as suggested by the BR
scaling \cite{br91}. What one does learn from the spacelike propagated
wave function \cite{KSBJ} is that the quarks in the $\rho$-meson behave, to
lowest order, as free quarks, each with thermal energy of $\pi T$,
with corrections of $\sim \alpha_s$ which give the Bethe-Salpeter
amplitude for the $\rho$. It is, in fact, this basically free quark
kinematics which leads to the equality $f_\pi^\star=f_s^\star$
for the spacelike wave functions. Wave functions propagated in the
time direction will, presumably, have this same basically free
quark kinematics. Indeed, in four-flavor lattice calculations of the
$\rho$-meson \cite{boyd2}, the temporal correlators above $T_{\chi SR}$
were found to be consistent with a pair of quarks that are
more or less free.  Modulo the caveat with four-flavor lattice calculations
mentioned before, given free quark kinematics, one
would expect $f_\pi^\star=f_s^\star$, even if $f_\pi^\star\rightarrow 0$
with chiral restoration. In the possible case of $f_\pi^\star\rightarrow 0$
as $T\rightarrow T_{\chi SR}$, the Georgi limit could be realized
either just before or simultaneously with the $f_\pi^\star$ going to zero.
Just what happens in the phase transition may be complicated to unravel.
Nonetheless, the lattice calculations of the susceptibility do show that,
to the accuracy we can analyze them, $g\rightarrow 0$ as $T\rightarrow
T_{\chi SR}$, indicative of the Georgi vector limit.

While other viable explanations may be found in the future, as far we know,
the soft kaons found in the E-814 experiment can be explained only if the
vector-meson coupling is essentially absent at the relevant temperature, and
this is a strong support for the Georgi picture in which the hidden gauge
coupling constant ``melts" at the temperature $T_{\chi SR}$.

Since Sakurai \cite{sakurai} we have known that vector dominance and universal
vector coupling worked well in describing vector interactions. It is in
endowing the vector mesons an induced gauge structure  \cite{bando} that
a contact (albeit an indirect one) with QCD is made. As pointed out by
Georgi \cite{vectorsym}, the hidden local symmetry is useful because it
keeps track of powers of $m_\rho/\Lambda_\chi$, where $\Lambda_\chi$ is the
scale of the effective chiral theory, in the situation where the
vector-meson mass $m_\rho$ is in some sense small. Indeed the successful
KSRF relation follows simply \cite{vectorsym} if the vector mesons as well
as the pseudoscalars are both light, so that we can stop with a Lagrangian
with the lowest (that is, two)
power of the derivative dictated by chiral invariance.
This suggests rather strongly that the notion that
the vector-meson mass becomes small as $T\rightarrow T_{\chi SR}$
is natural within the framework of effective chiral field theory.
Remarkably, this seems to be supported by lattice gauge calculations
at $T\sim T_{\chi SR}$. Now the gauge symmetry in the hidden symmetry
scheme is an ``induced" gauge symmetry lodged in hadronic variables.
In this sector, the fundamental color gauge symmetry is not
visible. It is the induced flavor one that is seen. What we observe is
then that as $T$ goes towards $T_{\chi SR}$, the induced gauge symmetry
gives way to the fundamental gauge symmetry. What is surprising is that
this changeover seems to take place suddenly, with increasing temperature,
the effective hadronic gauge symmetry applying for $T < T_{\chi SR}$,
and the fundamental color gauge symmetry being realized
perturbatively for $T > T_{\chi SR}$.

We have suggested that the Georgi vector symmetry with $\kappa=0$ and
$g\neq 0$  is also relevant
for nuclear physics. It is the existence of
such a symmetry that allows one to linearize the non-linear chiral
Lagrangian of current algebra to the linear sigma model which justifies
in some sense the separation of the scalar $\chi$ field,
the low-frequency part of which giving rise to the previously obtained
medium-scaling of effective chiral Lagrangians of ref.\cite{br91}.
Going beyond the
tree approximation with this scaled chiral Lagrangian in hot and/or
dense matter is an open problem which has attracted little attention
up to date. While consistent generally with both observations and
theoretical prejudices, our discussion cannot be considered solid
until higher-order calculations can be systematically carried out
and compared with experiments. A small progress made in this direction
is discussed in ref.\cite{mrelaf}.

Finally we should mention that an investigation of effective hadronic
interactions mediated by ``instanton molecules" carried out by Sch\"{a}fer
{\etal} \cite{schafer} provides support to our thesis that chiral
phase transition involves Georgi's vector limit. In this work, it is
found that all coupling constants in a NJL-type effective Lagrangian
can be specified for $T\gsim T_{\chi SR}$ in terms of a single
parameter $G$, signaling a swelling symmetry. At temperature
$T\gsim T_{\chi SR}$, the instantons can be taken to be completely
polarized. In this case, the interaction in the longitudinal vector-meson
channel becomes equally strong as the attraction in the scalar-pseudoscalar
channel. Transversely polarized vector mesons are found to have no interaction.
In detail how this comes about is as follows: The instanton molecules survive
the chiral symmetry restoring transition; they leave chiral symmetry unbroken.
According to Koch and Brown \cite{kochbrown}, about half of the gluon
condensate remains for temperature $T\gsim T_{\chi SR}$. This is assumed to
reside in the molecules \cite{schafer}. Quarks coupling through the instanton
molecules experience, for randomly oriented molecules, an interaction
\be
\delta H=\frac{2G}{N_c^2}\{ (\bar{\psi}\gamma_5\vec{\tau}\psi (x))^2
+\frac 14 (\bar{\psi}\gamma_\mu \vec{\tau}\psi (x))^2+\cdots\}
\ee
where we show only the terms that act in the pionic and $\rho$-meson
sector. The $\frac 14$ in front of the vector interaction comes from
averaging the molecules over random directions. In this case, it is
seen that the interaction in the $\rho$-meson channel is only $\frac 14$
as large as pionic channel. If the instanton is completely polarized in
the time direction -- and the authors of \cite{schafer} give reasons
why this lowers the energy of the system -- the interaction in the
$\rho$-meson channel changes to
\be
\frac 14 (\bar{\psi}\gamma_\mu\vec{\tau}\psi (x))^2\rightarrow
(\bar{\psi}\gamma_0\vec{\tau}\psi (x))^2
\ee
since directions are no longer averaged over. The $\frac 14$ factor no
longer appears, because there is no averaging over directions. This sets
interaction in the pionic channel and that for the time-like $\rho$ meson
equal, decoupling the transverse $\rho$ mesons. The time-like vector mesons
are longitudinal in the sense that
\be
q^\mu \rho_\mu=0
\ee
where $q_\mu$ is the four-momentum of the $\rho$ meson.
We thus have a ``realistic" dynamical model, giving  the effective interaction
in hadronic variables, which reproduces the Georgi vector limit under the
assumption of a dilute gas of fully polarized instanton molecules and,
of course, the $\rho$ meson going massless.\footnote{Sch\"{a}fer {\etal}
\cite{schafer} find that the molecules are not completely polarized (the
polarization being $\sim 70\%$), so this model may be taken only as a
caricature
theory which, in a limit not far away from the physical world,
reproduces the results of the Georgi vector limit.}
(In this model, the $\omega$
and the $\rho$ can be put into flavor $U(2)$ as well.) Note that the
above considerations apply to the region of $T$ just above $T_c$, where
the colored gluon exchange is present.

\section{Chiral Restoration in Dense Matter in Stellar Collapse:
Transition to Kaon Condensation and then to Quark Matter}
\indent

In Section 6 we discussed effects of the attractive kaon-nucleon
interaction, namely, the Kaplan-Nelson term, on subthreshold
production of $K^\pm$ mesons in heavy-ion collisions of Au on
Au, with 1 GeV/nucleon energy. These involve chiefly finite density
effects, because  the temperature is low, $\sim 75$ MeV, so that
$T$  is well below $T_c$. We do not expect much effect of dropping masses,
etc., because of the temperature.

Our consideration of chiral restoration with high temperature could be checked
by lattice gauge simulations; more correctly, many of our ideas about the
restoration stemmed from lattice gauge results. Unfortunately, lattice
calculations cannot be carried out for finite density -- at least not to date
-- because in changing to imaginary time (Euclidean space-time) the baryon
chemical potential becomes imaginary and the probability is no longer
positive, so that Monte Carlo methods no longer work, at least not effectively
in their present formulation.

Effects from finite density in strong interactions have been reviewed
extensively by Adami and Brown \cite{adamibrown} and so we will not repeat
this review here. A very interesting situation is however provided by
the collapse of stars, in which there is sufficient time for strangeness
conservation to be violated. We shall review recent developments on this
here.

For concreteness, let us sketch the scenario for Supernova 1987A.
According to the scenario we shall pursue here, the core of the
18 $M_\odot$ progenitor collapsed, initially forming a neutron star, and
blew off the outer material, in a time of $\sim 4$ seconds following collapse
\cite{bethe93}. During this time the neutron star was stabilized by the
trapped neutrinos. We know from the Kamiokande neutrino detector that
the neutrinos came off for $\sim 12$ seconds; by this time they had dropped
in energy below the $\sim 8$ MeV threshold for detectability.
According to Burrows and Lattimer \cite{burrows} only about half of the thermal
energy is carried away by the neutrinos during this $\sim 12$ seconds.
About half of the thermal energy remains and creates substantial pressure,
which helps to stabilize the neutron star. The central density of the
neutron star  grows to several times nuclear matter density.

Once the electron neutrinos have left, following the explosion, the
electrons in the core turn into $K^-$-mesons and neutrinos, as we now explain.
The neutrinos leave the core, again taking several seconds to do so.
This cannot happen while the first set of neutrinos are trapped, because this
would require
\be
\omega_{K} =\mu_K=\mu_e - \mu_\nu,
\ee
where the $\mu$'s are the chemical potentials, and as we shall see,
it would be difficult to bring the kaon energy $\omega_K$ down this far.
But when the $\mu_\nu\rightarrow 0$, it then becomes possible to bring
$\omega_K$ down to the electron chemical potential $\mu_e$.

Prakash, Ainsworth and Lattimer \cite{pal88} give simple parametrizations of
the neutron rich equation of state for the situation after the neutrinos have
left. The electron chemical potential is then established by the thermal
equilibrium condition
\be
e^- +p \leftrightarrow \nu + n,
\ee
with the neutrinos more or less  freely leaving the star, so
\be
\mu_e=\mu_n-\mu_p.
\ee
The chemical potentials of the neutron and proton must be calculated,
and $\mu_e$ determined from these and the condition of charge neutrality.
Although $\mu_e$ depends somewhat on the compression modulus, etc., a
good standard value for $\mu_e$ is
\be
\mu_e\cong 117 {\mbox MeV}\ \ \ {\mbox for}\ \ \rho=\rho_0.
\ee
The $\mu_e$ increases slightly faster than as $\rho^{1/3}$ with density,
reaching $\gsim 200$ MeV at about $3\rho_0$ for equations of state with a
nuclear matter compression modulus of $K_0=200$ MeV. These numbers
schematize the results of Thorsson, Prakash and Lattimer \cite{tpl94}
and of \cite{LBMR}.

Now the energy of a $K^-$-meson  has been found to drop in nuclear matter.
According to Friedman {\etal} \cite{friedmanetal}, the real part of the
$K^-$-nucleus potential is $-200\pm 20$ MeV in the center of the
$^{56}$Ni nucleus at $\rho=.97\rho_0$. The $K^-$ feels double the attraction
from a proton as it does from a neutron through the vector interaction
(\ref{repulsion}) and the same attraction for
neutron and proton through the scalar interaction (\ref{attraction}).
Thus, for neutron-rich matter, one can say that the $K^-$ energy is
decreased $\sim 150$ MeV at nuclear matter density $\rho_0$.
What can happen is schematically described in Fig.\ref{schematic}.

\begin{figure}
\vskip 5cm
\caption[scheme] {Behavior of the $K^-$ energy and of the electron
chemical potential $\mu_e$ as function of energy.}
\label{schematic}
\end{figure}

At the point where the $\omega_K$  line crosses the $\mu_e$ line,
it becomes energetically possible for the ``electrons" to turn into
$K^-$'s accompanied by neutrinos
\be
e^-\rightarrow K^- + \nu_e
\ee
with the neutrinos leaving the star. There will therefore be a second
burst of electron neutrinos which should be detectable.
Details (energy, time of emission, etc.) are being worked out by A. Burrows,
J.M. Lattimer, M. Prakash and collaborators in Stony Brook at the present
time.

We note that the large binding energies of Friedman, Gal and Batty
\cite{friedmanetal} fit in very well with the scenario of Brown-Rho scaling,
with $f\rightarrow f^\star$ in (\ref{repulsion}) and (\ref{attraction})
and as shown in Figs. 11-13 of the detailed calculations in \cite{LBMR}.
One seems to need as much attraction as the theory can provide.

At densities above the kaon condensation threshold, the $K^-$ mesons
of zero momentum form a Bose condensate, as discussed in
\cite{tpl94,LBMR}. Sufficient binding energy is gained so that the equation of
state is substantially softened, and the maximum neutron star mass is
$\sim 1.5 M_\odot$ as discussed by Brown and Bethe \cite{brownbethe}.
We will return to this matter at the end of this section.

It has been recently shown by Brown and Weingartner \cite{weingartner}
that in the case of Supernova 87A if a neutron star were present, then
we would see it with a luminosity $L$ of about $10^4$ times $L_\odot$,
whereas the bolometric luminosity actually is $L\approx 10^2 L_\odot$
and can be explained as arising from radioactivity in the expanding
shells. Thus their conclusion is that as in \cite{brownbethe} the core has
gone into a black hole. Furthermore Bethe and Brown \cite{bethebrown}
show that from the fact that 0.05 $M_\odot$ of $Fe$ was produced, together with
the pre-supernova evolution, it is possible to determine the mass of the
compact object to be in the range
(1.44 -- 1.56) $M_\odot$,
confirming the suggestion by Brown and Bethe \cite{brownbethe}.

Recently Keil and Janka \cite{keil} have shown, based on the Glendenning
mean-field equations \cite{glenn}, that hyperons enter into the equation of
state at densities $\rho \sim 2\rho_0$. They find a delayed explosion scenario
similar to that of Brown and Bethe \cite{brownbethe} for cores in the range
\be
1.58 M_\odot < M_{\mbox{grav}} <1.72 M_\odot.
\ee
Here we have used the binding energy of
\be
E=0.084 M_\odot (M/M_\odot)^2
\ee
of Lattimer and Yahil \cite{latya} in order to connect baryon number masses
given by Keil and Janka with gravitational masses. Undoubtedly parameters
could be adjusted to bring the lower limit down to the 1.5 $M_\odot$
of Bethe and Brown \cite{bethebrown}, which is required for observation.
More extensive calculations with inclusion of exchange (Fock) terms
\cite{huberetal} bring the maximum neutron star mass above 2 $M_\odot$,
the Fock terms significantly stiffening the EOS once hypersons are present.

Introduction of hyperons into the EOS in Hartree approximation causes
the electron chemical potential to saturate \cite{glenn} at $\mu_e
\sim 200$ MeV at $\rho\sim 2 \rho_0$. If hyperons enter at densities
before kaon condensation, they may hinder the condensation by bringing
$\mu_e$ down. On the other hand, they sufficiently soften the EOS that the
system moves rapidly to a higher density, which brings the $K^-$ energy
$\omega_K$ down, and these effects are likely to compensate.
As noted, linear extrapolation of the binding energy of a $K^-$ in nuclei
to higher densities, or inclusion of Brown-Rho scaling \cite{LBMR},
gives $\rho_c$ for kaon condensation only slightly above $2\rho_0$.
Once finite temperature is included, a substantial number of thermal
$K^-$-mesons are present, so the distinction between introduction of
hyperons and kaon condensation is likely to be smeared out.
Recently Ellis, Knorren and Prakash\cite{EKP} have introduced hyperons
into the dense matter, largely confirming the calculations of
Glendenning\cite{glenn} and of Keil and Janka\cite{keil}. Ellis {\etal}\
find that the introduction of $\Sigma^-$'s brings down the electron
chemical potential $\mu_e$ and delays the kaon condensation until higher
densities. We believe this to be incorrect for the following reasons:
(1) $\Sigma^-$'s do not experience the same attraction in nuclei as
the $\Lambda$'s do\cite{battyetal}, although this is assumed in the
quoted calculation; a possible explanation is that
QCD sum-rule calculations\cite{jiu} find substantially more repulsive
interaction on the $\Sigma^-$ than on the $\Lambda$, in contrast to
quark scaling which predicts these to be the same; (2) the Ellis
{\etal}\ calculations use Walecka mean fields for the hyperons, but
interactions
from chiral Lagrangians for the $K^-$ mean fields. As noted at the end
of Subsection 6.2, this is inconsistent. Given the Walecka mean fields
for hyperons, the $K^-$ mean fields should be a factor $\sim$ 1.6 larger
than employed. This not only makes kaon condensation competitive with
the introduction of hyperons, but also means that it will probably
precede any such introduction. To put it more directly,
we believe that kaon condensation will win out over introduction of
hyperons. Even if the latter did indeed enter first at lower densities,
it would soften the EOS sufficiently in order to lead quickly to kaon
condensation.

Aside from substantially softening the EOS, so that a compact object of
mass 1.5$M_\odot$ can go into a black hole, the strangeness condensed
EOS has an important consequence, shown in Fig.\ref{Y} \cite{gebfigure}.
While the neutrinos are trapped, the leptons (neutrinos and electrons)
produce substantial pressure, so that the maximum mass given by the solid
line in Fig.\ref{Y} is stabilized. Once the neutrinos leave, the
electron fraction $Y_l$ drops to a low value, and since most of the lepton
pressure has been carried away by neutrinos, the maximum mass that can
be stabilized is substantially less, 1.5$M_\odot$ in the case of
Fig.\ref{Y}. Consequently, compact objects with masses in the range
\be
1.5 M_\odot <M \lsim 1.7 M_\odot
\ee
will be stable during the time of neutrino emission; {\ie},\ long enough to
explode, and will then go into a black hole. Brown and Bethe \cite{brownbethe}
estimate that this delayed drop, after explosion and return of matter to the
galaxy, into a black hole will take place for most stars
with main sequence masses
\be
18 M_\odot < M < 30 M_\odot.
\ee

\begin{figure}
\vskip 10cm
\caption[geb] {The solid line shows the short-time maximum neutron star
masses as function of central density, with an assumed lepton fraction of
$Y_l=0.4$, consisting of electrons and trapped neutrinos. The dashed line
shows the long-time maximum cold mass, after the neutrinos
have left. Note that the highest $n_{cent}$ is $\sim 12\rho_0$.}
\label{Y}
\end{figure}

\begin{figure}
\vskip 15cm
\caption[EOS] {A very soft EOS with $K_0=130$ MeV so that the maximum
neutron star mass for cold matter (dashed line) is $1.5 M_\odot$ as in
Fig.\ref{Y}. The short time maximum mass, given by the solid line,
is now lower, even though the neutrinos are trapped with $Y_l=0.4$}
\label{Z}
\end{figure}

Supernova 87A lies near the lower end; in fact, it essentially sets the
lower end. In PSR 1913 $+$ 16 the pulsar, the larger of the neutron stars
in the  binary, has mass 1.44$M_\odot$. Evolutionary
calculations \cite{woosley,bww} find that a 1.44$M_\odot$ neutron star results
from a 5 -- 6 $M_\odot$ helium star, corresponding to a main sequence mass
of 16 -- 18 $M_\odot$ for the progenitor. Thus, the pulsar in 1913 + 16 is
very close to the maximum mass for a neutron star.

Not only do we believe that the Supernova 87A ended up as a black hole,
but this was also likely for CAS A, a supernova explosion that took place
in the 17th
century. From the abundances of O, Mg, Ne in knots in the supernova remnant,
the masses of the progenitor can be deduced to be $M\approx 20 M_\odot$.
No compact object is found in the center of the remnant, although searches,
with great sensitivity, have been carried out.

We shall now explain why the delayed explosion does not result from the
standard neutron star scenario. This situation is shown in Fig.\ref{Z}.
In this case the short-time maximum mass, with neutrinos trapped, and
with $Y_l=0.4$, is lower than the cold mass, given by the dashed line, after
the neutrinos leave. The point is that, in the standard scenario, as the
neutrinos leave, the original nuclear matter is converted to neutron matter,
and neutron matter is much ``stiffer" (that is, the pressure is higher)
than nuclear matter. Thus, even though the pressure is lowered by the
neutrinos departing, it is increased somewhat more, by the protons changing
to neutrons. Thus, if the compact object is initially stable, it will remain
stable, and will not go into a black hole.

Previous to the scenario involving the kaon condensed EOS, it was suggested
that stars heavier than $\sim 25 M_\odot$ may leave black holes \cite{wilson},
and also that such stars might first explode, exhibiting light curves of
Type II supernovae, and then collapse into black holes \cite{woosleyweaver}.
The compact core was, for certain range of masses, to be stabilized by the
thermal pressure during the period of Klein-Helmholtz contraction, long enough
to carry out nucleosynthesis, then going into a black hole after cooling and
deleptonization. In terms of Fig.\ref{Z}, this scenario meant that the thermal
pressure had to be sufficient, so that when thermal effects were added to
the (short-time) solid line, it came close to the dashed line. Then, the
compact object could have a mass lying above the dashed line (and below the
solid line) and be stable for some time, before it went below the dashed
line as it cooled.

     \begin{center}
     {\bf Table 1}\\
\vskip 0.3cm
     \parbox[t]{5.5in}{
     Energy gain $\Delta E$ in MeV, chemical potential $\mu_K$ in MeV,
proton fraction $x$ and electron fraction $x_e$ as function of the
density $u=\rho/\rho_0$ for a kaon condensed EOS with $u_c=4.2$.}
     \end{center}
    $$
    \begin{array}{|c||c|c|c|c|}
    \hline
    u & \Delta E ({\mbox MeV})& \mu_K ({\mbox MeV}) & x & x_e  \\ \hline \hline
       4.2 &  0 & 256 & 0.20 & 0.11  \\ \hline
       5.2 &  -10 & 199 &  0.34 & 0.03 \\ \hline
       6.2 &  -35 & 142 &  0.43 & 0.01 \\ \hline
       7.2 &  -71 &  94 &  0.48 & 0  \\ \hline
       8.2 & -112 &  55 &  0.50 & 0  \\ \hline
     \end{array}
     $$
\vskip 0.5cm

The trouble with this scenario, of thermal pressure stabilizing the compact
object, is that the thermal pressure can stabilize only a small additional
mass. Detailed calculations by Bombaci {\etal} \cite{bomba} give only an
additional $\sim 2$--$3\%$ increase in stabilized mass, less than the
distance between the solid and dashed lines in Fig.\ref{Z}. Similar
results are obtained by Keil and Janka \cite{keil} with their delayed explosion
scenario in the EOS including hyperons. Thus, it is clear that in most
cases, thermal pressure will not stabilize the compact object for
some time, with the object later going into a black hole; rather, it will
remain
stable if it is so initially.

We see that a lot of observational evidence is explained with our kaon
condensation EOS. Note that in this scenario, stars go ``strange"; {\ie},
\ acquire a lot of strangeness already in the hadron sector. They do not,
at the densities of 2 -- 4 $\rho_0$, go to strange quark matter.
The transition from kaon condensation to strange quark matter can, however,
be constructed \cite{brownquark}. This was made under quite conservative
assumptions of a small $\Sigma_{KN}=1.3 m_\pi$ and no Brown-Rho scaling.
We give in Table 1 the results of the calculation by Vesteinn Thorsson
in his 1992 Stony
Brook thesis. With small differences, these are the same numbers as in
Table 3 of ref.\cite{BLRT}.

Of chief concern to us is the softening of the EOS by kaon condensation.
In Fig.\ref{FIGA} we plot the baryon number chemical potential vs. pressure.
This is a convenient plot for investigating the transition to quark matter.
In the following we use the procedure of Bethe {\etal} \cite{BBC}.
The PAL curve is taken from Prakash {\etal} \cite{pal88}. It has
a compression modulus $K_0=180$ MeV, and the potential energy part of
the symmetry energy rises linearly with $u=\rho/\rho_0$.
The PAL21 curve from \cite{pal88} has a maximum neutron star mass of
$M_{max}=1.72 M_\odot$. The curve including the kaon condensation
with characteristics shown in Table 1 is plotted in the lower dashed
line in Fig.\ref{FIGA}. In addition, the results for quark matter \cite{BBC}
are given, but with strange quark mass $m_s=200$ MeV included.
(In \cite{BBC}, $m_s$ was set equal to zero.) There $\alpha_s$ was scaled
as
\be
\alpha_s (k_F)=2.2 k_F^{(0)}/k_F \label{MITscale}
\ee
where $k_F^{(0)}$ is the Fermi momentum at nuclear matter density,
$k_F$ the Fermi momentum at the density considered. This gave a rapid
decrease  in coupling constant, which Bethe {\etal} \cite{BBC}
considered appropriate for the nonperturbative sector.

\begin{figure}
\vskip 13cm
\caption[pal] {\small
Curves of chemical potential vs. pressure for hadronic matter
(dashed curves). The upper dashed curve representing PAL21\cite{pal88}
has $K_0=180$ MeV, a symmetry energy for which the potential energy rises
linearly with density. The lower dashed curve results when kaon condensation
is included. Solid lines for quark matter are plotted for $p_Q +B$, instead
of the pressure, but $B$ will be taken to be zero. The upper solid line is
for $\alpha_s=1.1$; the lower one, for $\alpha_s=0.55$. On each line,
the lowest black circle or square marks the density $2\rho_0$, and each
successive dot or square indicates a density higher by $\rho_0$.}
\label{FIGA}
\end{figure}

It should be noted that  the M.I.T. $\alpha_s$ of 2.2 at $\rho=\rho_0$,
even decreasing as rapidly as (\ref{MITscale}), gave an EOS which lies
well above the PAL21 curve, so there is no hope of joining PAL21
to it, especially if PAL21 is decreased even further by kaon condensation.
(A ``conventional" compression modulus of $K_0=210\pm 30$ MeV \cite{blaizot}
would give an only slightly higher curve than PAL21.)

Quark EOS's for $\alpha_s=1.1$ and 0.55 are also shown. The $P_Q + B$
is plotted for the quark/gluon phase, so that introduction of a bag constant
(which does not affect the chemical potential $\mu$) can be made by shifting
the $P_Q+B$ curve to the left by the amount $B$, so as to obtain $P_Q$.
As outlined in \cite{adamibrown}, we believe the bag constant, the
$B_{\chi SB}$ of eq.(\ref{MITB}), to go to zero at chiral restoration.
Note that the quark/gluon curve for $\alpha_s=1.1$ is easily made
equal to PAL21 over a wide range of pressure about $P_N\sim 400
{\mbox MeV/fm^3}$ in this way. Similarly, the quark/gluon curve with
$\alpha_s=0.55$ can be made equal to the PAL21 with kaon condensate
curve, in the region of pressures which correspond to densities of
$\sim 8$--$10\rho_0$ for hadronic matter. (The density is not continuous
at the transition.) We see immediately that with the M.I.T. value
for $\alpha_s$ used by Bethe {\etal} \cite{BBC}, there is no hope of
making a transition to quark matter; the quark matter EOS has a much
higher energy than the hadronic one, especially when kaon condensation
is introduced into the latter.

Fahri and Jaffe \cite{fahri} have taken the position that the $\alpha_s$
for the dense matter properties may be substantially smaller than 2.2,
and have investigated the range of $\alpha_s$ used in the two curves
of our figure. Given these small $\alpha_s$, it is possible to discuss
the transition to quark matter.

With our conservative choice of the kaon condensed EOS, without
Brown-Rho scaling, and the $\alpha_s$ of $0.55\frac{k_F^{(0)}}{k_F}$,
a smooth join to quark matter can be made at a density of $\sim 8$--$10
\rho_0$. Whereas $\alpha_s$ may not drop as rapidly as $k_F^{(0)}/k_F$,
an $\alpha_s$ of $\sim 0.25$ at $\rho\sim 10 \rho_0$ is not unreasonable,
although we have no way at present of determining a quantitative value
at such high densities. We can thus say that, with a rather conservative
kaon condensed EOS (with $\rho_c\approx 4 \rho_0$), there is the
possibility of a smooth cross-over transition to quark matter at the
upper end of the densities obtained in compact objects, say, $\rho\sim
10\rho_0$. With a softer EOS, obtained with Brown-Rho scaling, with kaon
condensation taking place at $\rho\lsim 3\rho_0$, the transition density
to quark matter would be substantially higher.

It is daring -- and perhaps foolhardy -- to extrapolate to densities
$\sim 10\rho_0$, but we believe the replacement of electrons by $K^-$-mesons
at high densities to be a new idea with a solid foundation, which
qualitatively changes the conceptual situation in compact star matter.
We may expect the transition to quark matter at high densities to
take place similarly to the transition with temperature to quark/gluon
plasma, as discussed in Section 4; namely, the transition will be smooth
and gradual. At the lower densities it will be more convenient -- and
cleverer -- to use hadronic variables, but at the higher densities the
quark language will become more efficient. As argued, we may expect
the light-quark vector-meson masses to go to zero, with increasing
density, and chiral symmetry to be realized in the Georgi vector limit
preceding chiral restoration.

Given the kaon condensation phase transition, our EOS is so soft that it
is difficult to stabilize stars of known masses. Indeed with the PAL21 EOS
including kaon condensation, $M_{max}=1.42 M_\odot$, although raising
$K_0$ to the more conventional value of 200 MeV allows us to obtain
$M_{max}=1.5 M_\odot$. A slightly higher $K_0$ will be needed if the kaon
condensed EOS with Brown-Rho scaling is used, since this EOS is
substantially softened by this.

\begin{figure}
\vskip 15cm
\caption[star] {Measured masses of 17 neutron stars from Arzoumanian
{\etal} \cite{arzo}, with the lower limit on the mass of Vela X-1 from
Van Kerkwijk {\etal} \cite{vankerk}. Objects in high mass X-ray
binaries are at the top, radio pulsars and their companions at the
bottom.}
\label{FIGB}
\end{figure}

Observed neutron star masses are shown in Fig.\ref{FIGB}. Precisely the
lower limit of the measured mass of Vela X-1 lay below $1.5 M_\odot$.
New observation by Van Kerkwijk {\etal} \cite{vankerk} found that  the
observed velocities in Vela X-1 deviate substantially from the smooth
radius-velocity curve expected from pure Keplerian motion. The deviations
seem to be correlated with each other within one night, but not
from one night to another. The excursions suggest something like
pulsational coupling to the radial motion, and make it difficult to
obtain an accurate mass measurement. The lower limit for the mass of the
compact object in Vela X-1 is now found to be $1.43 M_\odot$ at 95\% confidence
level, or $1.37 M_\odot$ at this confidence interval around the most
probable value. Consequently, Vela X-1 is no longer a big
problem for our $M_{max}=1.5 M_\odot$.

It is striking that well measured neutron star masses lie below
$1.5 M_\odot$. (See Fig.\ref{FIGB}.) However, the central value
of the compact object in 4U 1700-37 lies at $1.8 M_\odot$,
although the error bars encompass $1.5 M_\odot$. Brown {\etal} \cite{bww}
give arguments that this compact object could be a low-mass black hole.
(The principal aim of this paper is to show that stars in binaries can
have much larger masses than the main sequence mass of $18 M_\odot$ of
1987A, and still end up as neutron stars. In other words, in binaries,
because of the specifics of mass transfer, stars can evolve in quite a
different way than single stars evolve. Since determination of masses has
generally been carried out in binaries, this explains why
it was not earlier recognized that a star with main sequence mass as light
as $\sim 18 M_\odot$ (the progenitor of 1967A) could go into a
black hole, and it was -- and still is -- considerable surprise that it
probably did.)

In the past it has generally been thought that the reason accurately
measured neutron star masses lie at $1.44 M_\odot$ and below is evolutionary
in nature. Large stars collapse when the iron core exceeds the Chandrasekhar
limit, $\sim 1.25$--$1.5 M_\odot$, depending on the main sequence mass of the
star. In the past literature, accretion has been assumed to proceed only
up to the Eddington limit
\be
\dot{M}_{Edd}=1.5\times 10^{-8} M_\odot/yr
\ee
and there are relatively few situations where the compact core would be
expected to accrete more than $\sim 0.1 M_\odot$
at this rate. (Even in the very old millisecond pulsars, only
$\sim 0.1$--$0.2 M_\odot$ is estimated to have been accreted.)
However, Chevalier \cite{chevalier} and Brown \cite{geb95}
have shown that in the common envelope phase of binary pulsar evolution,
accretion can proceed at {\it hypercritical rates}
\be
\dot{M}\geq 10^4 \dot{M}_{Edd}.
\ee
Thus, if evolutionary history determines their mass, neutron stars of
higher mass than $1.5 M_\odot$ should exist, but none have been so
far observed. Therefore, there should be an intrinsic limit on the mass
of a neutron star, such as the one we find. It is difficult to explain
the existence of such an intrinsic limit without softening of the EOS
through a phase transition and we are proposing that kaon condensation
is the key mechanism for it.

\subsection*{Acknowledgments}
\indent

We are grateful to Tetsuo Hatsuda, Volker Koch, Maciej Nowak,
Koichi Yamawaki and Ismail Zahed for helpful discussions and to
Madappa Prakash for advice and results on dense matter. We also thank
Chang-Hwan Lee for checking
some of the equations and making suggestions for improvement.
This paper was completed at the Institute for Nuclear Theory,
University of Washington, Seattle while one of the authors (MR)
was participating in the INT95-1 program on ``Chiral dynamics in hadrons
and nuclei." MR would like to thank the INT for its hospitality and the
Department of Energy for partical support.
\vskip 2cm

\subsection{Appendix: The Kaplan-Nelson Attraction}
\indent

The attractive scalar mean field potential (\ref{attraction}), used
first by Kaplan and Nelson\cite{kaplanelson} for kaon condensation,
depends linearly on the $KN$ sigma term $\Sigma_{KN}$, which in turn depends
on the strangeness content of the nucleon $\la N|\bar{s}s| N\ra$.
This is usually parametrized by
\be
y=\frac{2\la N|\bar{s}s|N\ra}{\la N|(\bar{u}u +\bar{d}d)|N\ra}.
\ee
In the most extensive lattice gauge calculation to date, Liu
finds \cite{liu}
\be
y=0.33\pm 0.09.
\ee
Given this, we can estimate $\Sigma_{KN}$ using the information on the
$\pi N$ sigma term $\Sigma_{\pi N}$.
Since
\be
\frac{\Sigma_{KN}}{\Sigma_{\pi N}}=\frac{(m_s+m_u)\la N|(\bar{s}s +\bar{u}u)|
N\ra)}{(m_u+m_d)(\la N|(\bar{u}u +\bar{d}d)|N\ra)},
\ee
taking the value for $m_s$ of
\be
2m_s/(m_u+m_d)\approx 29
\ee
from Bijnens {\etal} \cite{bijnens}, $y=0.33$ and $\Sigma_{\pi N}=45$ MeV,
we find
\be
\Sigma_{KN}\cong 450\pm 30\ \ {\mbox MeV}\label{bestest}
\ee
as the best current estimate of $\Sigma_{KN}$. This is slightly larger
than the 2.83 $m_\pi$ found by Lee {\etal} \cite{LBMR} from fitting the
$KN$ scattering amplitudes. We believe that the wild fluctuations in
$\Sigma_{KN}$ used in the literature have now settled down.

Using Liu's value\cite{liu} of
\be
\la N|\bar{u}u+\bar{d}d|N\ra=8.22\pm 1.1,
\ee
we find as central value
\be
\la N|\bar{s}s|N\ra=1.36.
\ee
With $m_s=174$ MeV, this would mean that $m_s\la N|\bar{s}s|N\ra$, the
contribution to the nucleon mass from the explicit chiral symmetry
breaking in the strange sector, is 237 MeV which is sizable.

In chiral Lagrangians, the explicit chiral symmetry breaking in the strange
sector is parametrized by a coefficient denoted $a_3$ in \cite{politzer}.
Politzer and
Wise employed rather different values, $a_3 m_s=310$ MeV corresponding to
a large $\la N|\bar{s}s| N\ra$ and $a_3 m_s=140$ MeV corresponding to a
small $\la N|\bar{s}s|N\ra$\footnote{Note that we are using a different
sign convention.}. Using our above central values for $m_s$ and $\la
N|\bar{s}s|N\ra$, we find, using the relations given by Politzer and Wise,
\be
a_3 m_s\approx 245 \ {\mbox{MeV}}
\ee
with an estimated uncertainty of $\sim 10\%$.

Note that the $\Sigma_{KN}=450$ MeV of eq.(\ref{bestest}) would give a
coefficient of 72 MeV, rather than 64 MeV, in eq.(\ref{attraction}).
This difference is not significant, given the listed uncertainties.

\end{document}